\documentclass[twocolumn, eqsecnum, prd, superscriptaddress]{revtex4-2}
\usepackage{amsmath,amssymb}
\usepackage{amsthm}
\usepackage{appendix}
\usepackage{comment}
\usepackage[dvipdfmx]{graphicx}
\usepackage[colorlinks]{hyperref}
\usepackage{grffile}
\usepackage{mathrsfs}
\usepackage{mathtools}
\usepackage{bm}
\usepackage{url}
\usepackage[usenames]{color}
\usepackage{ulem}
\usepackage{natbib}
\usepackage{comment}
\usepackage{xspace}

\def\O{\mathcal{O}}

\def\gwh{gravitational-wave\xspace}

\newcommand{\RE}[1]{\mathrm{Re} \left[ {#1} \right]}
\newcommand{\IM}[1]{\mathrm{Im} \left[ {#1} \right]}
\newcommand{\Sub}[2]{{#1}_\mathrm{#2}}
\newcommand{\Super}[2]{{#1}^\mathrm{#2}}
\newcommand{\intd}{\mathrm{d}}

\newcommand{\pdiff}[2]{\frac{\partial {#1}}{\partial {#2}}}
\newcommand{\ve}[1]{\bm{{#1}}}
\newcommand{\SupSub}[3]{{#1}^\mathrm{#2}_\mathrm{#3}}

\newcommand{\pmat}[1]{\begin{pmatrix}#1\end{pmatrix}}

\begin{document}
\title{Assessing the impact of non-Gaussian noise on convolutional neural networks that search for continuous gravitational waves}
\author{Takahiro S. Yamamoto}
\email{yamamoto.takahiro.u6@f.mail.nagoya-u.ac.jp}
\affiliation{Department of Physics and Astrophysics, Nagoya University, Nagoya 464-8602, Japan}
\author{Andrew L. Miller}
\email{andrew.miller@uclouvain.be}
\affiliation{Universit\'{e} catholique de Louvain, Chemin du Cyclotron 2, B-1348 Louvain-la-Neuve, Belgium}
\author{Magdalena Sieniawska}
\affiliation{Universit\'{e} catholique de Louvain, Chemin du Cyclotron 2, B-1348 Louvain-la-Neuve, Belgium}
\author{Takahiro Tanaka}
\affiliation{Graduate School of Science, Kyoto University, Kyoto 606-8502, Japan}
\affiliation{Center for Gravitational Physics, Yukawa Institute for Theoretical Physics, Kyoto University, Kyoto 606-8502, Japan}
\date{\today}
\begin{abstract}
    We present a convolutional neural network that is capable of searching for continuous gravitational waves, quasi-monochromatic, persistent signals arising from asymmetrically rotating neutron stars, in $\sim 1$ year of simulated data that is plagued by non-stationary, narrow-band disturbances, i.e., \textit{lines}. Our network has learned to classify the input strain data into four categories: (1) only Gaussian noise, (2) an astrophysical signal injected into Gaussian noise, (3) a line embedded in Gaussian noise, and (4) an astrophysical signal contaminated by both Gaussian noise and line noise. In our algorithm, different frequencies are treated independently; therefore, our network is robust against sets of evenly-spaced lines, i.e., \textit{combs}, and we only need to consider perfectly sinusoidal line in this work. We find that our neural network can distinguish between astrophysical signals and lines with high accuracy. In a frequency band without line noise, the sensitivity depth of our network is about $\mathcal{D}^{95\%} \simeq 43.9$ with a false alarm probability of $\sim 0.5\%$, while in the presence of line noise, we can maintain a false alarm probability of $\sim 10\%$ and achieve $\mathcal{D}^\mathrm{95\%} \simeq 3.62$ when the line noise amplitude is $h_0^\mathrm{line}/\sqrt{S_\mathrm{n}(f_k)} = 1.0$. The network is robust against the time derivative of the frequency, $\dot{f}$ of a \gwh signal, i.e., the spin-down, and can handle $|\dot{f}|\lesssim 10^{-12}$ Hz/s, even though our training sets only included signals with $\dot{f}=0$. We evaluate the computational cost of our method to be $\mathcal{O}(10^{19})$ floating point operations, and compare it to those from standard all-sky searches, putting aside differences between covered parameter spaces. Our results show that our method is more efficient by one or two orders of magnitude than standard searches. Although our neural network takes about $\O(10^8)$ sec to employ using our current facilities (a single GPU of GTX1080Ti), we expect that it can be reduced to an acceptable level by utilizing a larger number of improved GPUs.
\end{abstract}
\maketitle

\section{Introduction}

Gravitational waves that are characterized by quite long durations, quasi-monochromatic frequencies, and almost constant amplitudes are called \textit{continuous gravitational waves} (CGWs) (see \cite{Lasky:2015uia, Glampedakis:2017nqy, riles2017recent,Sieniawska:2019hmd,Tenorio:2021wmz} for reviews). In the source frame, the waveform model of a CGW is governed by a small number of parameters: an amplitude, an initial frequency, and a first (and higher) time derivative(s) of the frequency evolution.

Several sources are expected to emit CGWs, the most promising of which are distorted, rotating neutron stars. In radio astronomy, rotating neutron stars are already observed as pulsars~\cite{2004hpa..book.....L}, whose rotational frequencies and sky locations are accurately estimated, according to the ATNF (Australia Telescope National Facility) Pulsar Database~\footnote{\url{http://www.atnf.csiro.au/people/pulsar/psrcat/}} ~\cite{Manchester:2004bp}. Therefore, this information can be used to look for CGWs emitted from these pulsars. This type of search is classified as a \textit{targeted search}, because the source rotational frequency, its derivatives, and the sky position are known, which allows us to perform a deep, fully coherent analysis for CGWs. However, these searches are limited to $\mathcal{O}(100)$ known pulsars; thus, other types of analyses are needed for objects about which we have less information.

Supernova remnants could also be interesting sources of CGWs, since they could house a compact object, such as a neutron star, at its center. Although the rotation frequency of the central object is unknown, the source location can be (roughly) identified. Therefore, we can use \textit{directed search} methods that do not assume a rotational frequency, which unfortunately requires a higher computational cost than that in targeted searches. Semi-coherent methods had to be designed to make a search for these objects tractable, which reduces their sensitivity in comparison to that in targeted searches; however, the exciting possibility of discovering CGWs from something unknown motivates looking at such systems. 

The most difficult, computationally heavy search is called a \textit{blind search} or \textit{all-sky search}, in which we do not have any information about potential CGW sources, e.g. electromagnetically silent neutron stars, ultra-light boson clouds around rotating black holes~\cite{Arvanitaki:2009fg,Brito:2015oca,DAntonio:2018sff,isi2019directed,Palomba:2019vxe}, inspiraling binaries consisting of planetary-mass black holes~\cite{Miller:2020kmv,Miller:2021knj, Pujolas:2021yaw} or of an ordinary compact object and a much light exotic one with an extreme mass ratio \cite{Guo:2022sdd}. The computational difficulties in all-sky searches arise because we would observe CGWs in a moving detector frame. Thus, the observed phase of a CGW is modulated by Doppler effects due to the relative motion between the source and the detectors. And since we do not know where in the sky CGWs could come from, we must search each location individually, over all possible source parameters. Despite the computational cost, many clever methods (e.g. Time-domain $\mathcal{F}$-statistic~\cite{Jaranowski:1998qm}, Frequency Hough~\cite{Astone:2014esa}, Sky Hough~\cite{Krishnan:2004sv}, and Hidden Markov Model~\cite{Bayley:2019bcb}) have performed all-sky searches, resulting in competitive constraints on the amplitude of CGWs over the whole sky ~\cite{LIGOScientific:2017csd,LIGOScientific:2017wva,LIGOScientific:2018gpj,LIGOScientific:2019yhl,LIGOScientific:2020qhb,LIGOScientific:2021inr,LIGOScientific:2021jlr,LIGOScientific:2021quq,LIGOScientific:2022pjk}. 

Another difficulty in CGW searches comes from the detector's non-Gaussian artifacts, which is known as $\textit{line noise}$~\cite{LSC:2018vzm}. Line noise has an almost constant frequency and a much larger amplitude than the Gaussian component of the detector noise. Because of line noise, standard methods for all-sky searches are required to veto frequency bands where line noise is present. This makes the analysis blind around those frequencies since these lines are often spread into multiple frequency bins and have multiple harmonics. It is, therefore, necessary to devise algorithms that are not only computationally efficient but also robust against such artificial disturbances.

In the last five years, the application of deep learning has been widely discussed in gravitational-wave astronomy (see~\cite{Cuoco:2020ogp} for review), and there are several proposals on the application to the detection and parameter estimation of various sources~\cite{George:2016hay, George:2017pmj, Colgan:2019lyo, Schmidt:2020yuu, Gabbard:2019rde, Chua:2019wwt, Green:2020hst, Dax:2021tsq, Kuo:2021qtt, Nakano:2018vay, Shen:2019vep, Yamamoto:2020rse, Bhagwat:2021kfa, Miller:2019jtp, Miller:2018rbg,Morras:2021atg,Chatterjee:2019avs,mytidis2015sensitivity}.
Deep learning algorithms could provide a way of alleviating both high computational costs and extreme sensitivity to noise lines in all-sky searches. As we will show, neural networks could be trained on different noise artifacts, allowing a systemic discrimination between them and astrophysical signals. Furthermore, after training, neural networks can generally classify new data in different categories very quickly. Such methods could therefore be used alongside existing ones, allowing the standard searches to be performed with increased sensitivity.

In particular, for CGW searches, several groups have proposed deep learning to analyze long stretches of strain data with durations of $O(10^{5-7})$ sec, all of which treat the data differently before feeding it to the neural network. For example, Dreissigacker \textit{et al.} \cite{Dreissigacker:2019edy} use the Fourier transformation to preprocess the data. They prepare several neural networks trained with the dataset corresponding to different frequency bands. Although the effects of non-Gaussian noise were not considered, they showed that their sensitivity is comparable to that of the semi-coherent matched filter. 

Combining deep learning with an existing analysis method is also a possible direction of research. Morawski \textit{et al.} \cite{Morawski:2019awi} employed the time domain $\mathcal{F}$-statistic for each grid point in the parameter space as inputs to their neural network. Their network was constructed to classify the strain data into three classes; only Gaussian noise, CGW signal in Gaussian noise, and sinusoidal line with Gaussian noise. For data with a duration of two days, their method discriminated the aforementioned three cases with high accuracy but did not handle the case in which line noise and CGWs exist in the same data stream. 

Beheshtipour and Papa~\cite{Beheshtipour:2020zhb} applied neural networks in the follow-up stages of the Einstein@Home pipeline in order to identify clusters of interesting candidates within the parameter space. They reported a slightly improved sensitivity compared to the results of an all-sky search in LIGO/Virgo's first observing run \cite{LIGOScientific:2017wva}; however, the computational cost was not reduced because Einstein@Home already requires a significant amount of computing power to perform the deepest all-sky searches in the CGW community.

Bayley \textit{et al.}~\cite{Bayley:2020zfa} combined deep learning and the Viterbi algorithm proposed in~\cite{Bayley:2019bcb}, and analyzed data with mock signal injections from the sixth science run of initial LIGO/Virgo \cite{Walsh:2016hyc}. They found that their method achieves comparable sensitivity to semi-coherent searches at a much lower computational cost. However, their neural networks specialized in detection and did not predict the source location.

Our previous work~\cite{Yamamoto:2020pus} demonstrates that a convolutional neural network can detect CGWs in a single detector output that contains stationary Gaussian noise. We proposed a new preprocessing method in which a double Fourier transform is applied to strain data that is partially demodulated by the time resampling. This preprocessing step concentrates the power of CGW signals into a small number of data points. The signal strength is therefore enhanced so that the neural network can easily detect CGWs. Our neural network independently treats the data of different source locations and frequency bins. Therefore, the computational cost of the follow-up can be reduced by specifying the parameter region where the follow-up is carried out. Although the sensitivity seems much better than that of the existing coherent search, it is demonstrated under the assumption that CGWs are circularly polarized (i.e., $\cos\iota = 1$ and $\psi=0$). We no longer use this assumption in this work.

While there have been many efforts to use machine learning to detect CGWs, none of them systemically address the problem that non-Gaussian noise pollutes real GW data containing astrophysical signals. Even those that have been applied to real data do not provide a recipe to handle non-Gaussian noise, nor do they indicate concretely how different line noise strengths affect their sensitivity and false alarm probability. The existing literature is therefore not systematic enough to be easily applied to future observing runs.

In this paper, we extend the work of \cite{Yamamoto:2020pus} to the case in which the strain data is contaminated by line noise. Our current study demonstrates that our network can handle more realistic GW data, and that the preprocessing step enables the network to be robust against line noise. Furthermore, our method in which different frequencies are treated independently would be robust against ``combs'', that is, a bunch of evenly-spaced lines. We also show how the sensitivity degrades with increasing line noise strength, in comparison to that obtained in Gaussian noise, and how high spin-downs of a CGW could affect the sensitivity of our network. This paper demonstrates that neural networks must consider the impact of non-stationary noise in CGW searches, and provides a more realistic comparison of the performance of neural networks to other all-sky methods.

We organize the rest of the paper as follows: in Sec.~\ref{sec:waveform and preprocess}, we describe a waveform and line noise model, and how we process the strain data before feeding it into a convolutional neural network. In Sec.~\ref{sec: method}, we explain our strategy to search for CGWs using a convolutional neural network. We show in Sec.~\ref{sec: results} sensitivity and false alarm probability estimations in the presence of Gaussian noise and Gaussian noise polluted by line noise, as well as robustness against small signal frequency changes. Following that, we estimate the computational cost of this method in Sec.~\ref{sec:computational cost}, and we make some concluding remarks and discuss ideas for future work in Sec.~\ref{sec:conclusion}.

\section{Waveform models and preprocessing}
\label{sec:waveform and preprocess}

In this work, we assume that the spectral density of Gaussian noise is stationary. The total duration and the sampling frequency are denoted by $\Sub{T}{dur}$ and $\Sub{f}{s}$, respectively. We fix them as $\Sub{T}{dur} = 2^{24}$ sec ($\sim 192$ days) and $\Sub{f}{s} = 1024$ Hz. See also Table~\ref{tab:preprocess parameters}, which shows the parameters of the strain data and the preprocessing step.

\begin{table*}[t]
    \centering
    \caption{\label{tab:preprocess parameters}
    List of parameters characterizing the strain data and the preprocessing.}
    \begin{ruledtabular}
    \begin{tabular}{lll}
        Description & Symbol & Value \\ \hline
        Sampling frequency of the strain data & $\Sub{f}{s}$ & 1024 [Hz] \\
        Total duration of the strain data & $\Sub{T}{dur}$ & 16777216 [sec] \\
        Threshold of the phase modulation after the time resampling & $\delta\Phi_\ast$ & 0.01 \\
        The number of grid points & $\Sub{N}{grid}$ & 5609178 \\
        Duration of SFT segment & $\Sub{T}{seg}$ & 2048 [sec] \\
        The number of data points within a SFT segment & $L$ & 2097152 \\
        Steepness parameter of Tukey window & $\xi$ & 0.125 \\
        The number of SFT segments & $\Sub{N}{seg} (=\Sub{T}{dur}/\Sub{T}{seg})$ & 8192 \\
        Upper limit of frequency which we analyze & $\Sub{f}{up}$ & 100 [Hz]\\
        The number of frequency bins & $\Sub{N}{bin} (= \Sub{T}{seg}\Sub{f}{up})$ & 204800
    \end{tabular}
    \end{ruledtabular}
\end{table*}

\subsection{Astrophysical signal}
The observed signal depends on the antenna pattern and the phase evolution. The waveform that we observe can be written as~\cite{Jaranowski:2009zz}
\begin{align}
    h_\mathrm{obs}(t) \coloneqq h_0 &\bigg[ F_+(t) \frac{1+\cos^2\iota}{2} \cos \Phi(t) \notag\\
    &\quad + F_\times (t) \cos\iota \sin \Phi(t) \bigg] \,.
    \label{eq: signal model}
\end{align}
where $h_0$ is the amplitude of the signal, $\iota$ is the inclination angle, and $F_+(t)$ and $F_\times(t)$ are the antenna pattern functions that depend on the source's location on the sky, the geometric configuration of the interferometer, and the location of the detector on Earth. The definitions of $F_+(t)$ and $F_\times(t)$ are the same as those used in Jaranowski \textit{et al.} \cite{Jaranowski:1998qm}, and we assume a LIGO-Hanford detector~\cite{LIGOScientific:2014pky}. $\Phi(t)$ is the observed phase of the gravitational waves, which we model to include the Doppler effect, the frequency $\Sub{f}{gw}$ and the first time derivative of the frequency $\dot{f}$
\begin{equation}
  \Phi(t) = 2\pi \Sub{f}{gw} \left( t + \frac{\ve{r}(t) \cdot \ve{n}}{c} \right) + \pi \dot{f} \left( t + \frac{\ve{r}(t) \cdot \ve{n}}{c} \right)^2 + \phi_0\,,
  \label{eq: signal phase}
\end{equation}
where $\phi_0$ is the initial phase, and $\ve{n}$ is the unit vector pointing to the source:
\begin{equation}
    \ve{n}(\alpha, \delta) =
    \left(\begin{matrix}
    1 & 0 & 0 \\
    0 & \cos\epsilon & \sin\epsilon \\
    0 & -\sin\epsilon & \cos\epsilon 
    \end{matrix}\right)
    \left(\begin{matrix}
    \cos\alpha\cos\delta \\ \sin\alpha\cos\delta \\ \sin\delta
    \end{matrix}\right)\,.
\label{sourcedirection}
\end{equation}
Here, $\alpha$ is the right ascension, $\delta$ is the declination, and $\epsilon$ is the tilt angle between the Earth's rotation axis and the orbital angular momentum. Here, we set the $x$-axis to point towards the vernal equinox and the $z$-axis to be along the Earth's orbital angular momentum. We assume that the position vector of the detector $\ve{r}(t)$ can be decomposed into the Earth's rotation, $\ve{r}_\oplus (t)$, and the Earth's orbital motion, $\ve{r}_\odot (t)$, i.e.,
\begin{equation}
    \ve{r}(t) = \ve{r}_\odot(t) + \ve{r}_\oplus(t)\,.
\end{equation}
We neglect various effects, such as the orbital eccentricity and the influence of the other planets and moon, and assume that the Earth follows a circular orbit on the $xy$-plane. These effects would be taken into account properly by using more sophisticated ephemeris in the preprocess stage. We write the orbital motion of the Earth as
\begin{equation}
\ve{r}_\odot(t) = R_\mathrm{ES}
\left(\begin{matrix}
\cos (\varphi_\odot + \Omega_\odot t) \\ \sin (\varphi_\odot + \Omega_\odot t) \\
0
\end{matrix}\right)\,,
\label{solarmotion}
\end{equation}
where $R_\mathrm{ES}$, $\Omega_\odot$ and  $\varphi_\odot$ are 
the distance between the Earth and the Sun, the angular velocity of the orbital motion and the initial phase, respectively.
The detector motion due to the Earth's rotation is
\begin{equation}
\ve{r}_\oplus(t)
=R_\mathrm{E}
\left(\begin{matrix}
1 & 0 & 0 \\
0 & \cos\epsilon & \sin\epsilon \\
0 & -\sin\epsilon & \cos\epsilon 
\end{matrix}\right)
\left(\begin{matrix}
\cos\lambda\cos (\varphi_\oplus + \Omega_\oplus t) \\ \cos\lambda\sin(\varphi_\oplus + \Omega_\oplus t) \\ \sin\lambda
\end{matrix}\right)\,,
\label{eq:detloc}
\end{equation}
where $R_\mathrm{E}$, $\lambda$, $\Omega_\oplus$ and $\varphi_\oplus$ are the radius of the Earth, the latitude of the detector, the angular velocity of the Earth's rotation and the initial phase, respectively. In this work, we fix $\varphi_\odot = \varphi_\oplus = 0$ for simplicity.

\subsection{Preprocessing}

It is known that preprocessing the data plays a crucial role in improving a neural network's performance~\cite{Goodfellow-et-al-2016}. We employ the method proposed in the previous work~\cite{Yamamoto:2020pus}, which we briefly review in this subsection.

Our method consists of three steps. In the first step, the strain data are transformed by a time resampling procedure. We prepare grid points on the sky, and define a new time coordinate for each grid point as
\begin{equation}
    \tau(t; \alpha_a, \delta_a) := t + \frac{\ve{r}(t) \cdot \ve{n}_a}{c}\,.
    \label{eq: new time coordinate tau}
\end{equation}
Here, $a$ is the index specifying the grid point and runs from 1 to $\Sub{N}{grid}$, the number of grid points on the sky. $\alpha_a$ and $\delta_a$ are the right ascension and the declination angle of the $a$-th grid point, respectively, and $\ve{n}_a \coloneqq \ve{n}(\alpha_a, \delta_a)$ is the unit vector pointing to the $a$-th grid point on the sky. Throughout the paper, the new time coordinate in Eq.~\eqref{eq: new time coordinate tau} is abbreviated as $\tau_a$. 
The resampled strain data is denoted by
\begin{equation}
  s_a(\tau) \coloneqq s(t_a (\tau))\,,
\end{equation}
where $s(t)$ is the strain data and $t_a(\tau)$ satisfies the relation
\begin{equation}
    \tau = t_a(\tau) + \frac{\ve{r}(t_a(\tau)) \cdot \ve{n}_a}{c}\,.
\end{equation}
The time grid is resampled such that in the new coordinate $\tau_a$, the grid is uniformly spaced.
Then, the phase modulation due to an astrophysical signal is removed, which results in signal power accumulating in a small number of frequency bins. Moreover, in the new time coordinate, sinusoidal line noise (see Sec.~\ref{sec: line noise}) is no longer monochromatic because it becomes Doppler modulated. That is why we expect an astrophysical signal can be discriminated from line noise.

The residual phase can be written as
\begin{equation}
  \delta\Phi_a(t) = 2\pi \Sub{f}{gw} \frac{\ve{r}(t) \cdot \Delta \ve{n}_a}{c}\,,
\end{equation}
where
\begin{equation}
  \Delta\ve{n} \coloneqq \ve{n} - \ve{n}_a\,,
\end{equation}
is the deviation between the $a$-th sky grid point and the gravitational-wave source.
We split the residual phase into two parts: the Earth's rotation:
\begin{equation}
    \delta\Phi^\oplus_a(t) \coloneqq 2\pi \Sub{f}{gw} \frac{\ve{r}_\oplus(t) \cdot \Delta \ve{n}_a}{c}\,,
\end{equation}
and the Earth's orbital motion:
\begin{equation}
    \delta\Phi^\odot_a(t) \coloneqq 2\pi \Sub{f}{gw} \frac{\ve{r}_\odot(t) \cdot \Delta \ve{n}_a}{c}\,.
\end{equation}
The grid points are placed on the sky such that the residual phase $\delta\Phi^\oplus_a(t)$ is suppressed below a threshold for any location of the source. 
This condition can be written as
\begin{equation}
  \min_a \max_t |\delta\Phi^\oplus_a (t)| \leq \delta\Phi_\ast\ \text{for any source}\,,
  \label{eq:condition of grid points}
\end{equation}
where $\delta\Phi_\ast$ is a threshold.
As in the previous work, we use the template placement method proposed in Nakano \textit{et al.}~\cite{Nakano:2003ma} to efficiently place grid points in a two-dimensional parameter space. First, we assume that the difference between the source direction and the grid point is small, and denote this difference by
\begin{equation}
    \Delta \delta_a \coloneqq \delta - \delta_a\,,\qquad
    \Delta \alpha_a \coloneqq \alpha - \alpha_a\,.
\end{equation}
We expand the residual phase $\delta\Phi_a^\oplus$ to the first order of $\Delta\alpha$ and $\Delta\delta$ as
\begin{align}
    \delta\Phi_a^\oplus(t) \simeq &\frac{2\pi \Sub{f}{gw}}{c} \Sub{R}{E} \cos\lambda \{ -\Delta\delta_a \sin\delta_a \cos(\alpha_a - \varphi_\oplus - \Omega_\oplus t) \notag\\
    &-\Delta\alpha_a \cos\delta_a \sin(\alpha_a - \varphi_\oplus - \Omega_\oplus t)\}\,,
\end{align}
while neglecting the constant term. The maximum of $\delta\Phi_a^\oplus(t)$ is 
\begin{align}
    \max_t |\delta\Phi_a^\oplus(t)| = &\frac{2\pi \Sub{f}{gw}}{c} \Sub{R}{E} \cos\lambda \notag\\
    &\times \sqrt{(\Delta\delta_a)^2 \sin^2\delta_a + (\Delta\alpha_a)^2 \cos^2\delta_a}\,.
\end{align}
Allowing $\cos\lambda$ to be 1 makes the estimation of $\max_t|\delta\Phi_a^\oplus(t)|$ conservative. Therefore, we use $\cos\lambda = 1$ in the following. We rewrite the condition~\eqref{eq:condition of grid points} as
\begin{equation}
    \min_a \left[ \Delta\sigma^2_a \right] \leq \delta\Phi_\ast^2 \left( \frac{c}{2\pi \Sub{f}{gw} \Sub{R}{E}} \right)^2 \text{ for any source}\,,
    \label{eq:condition of grid points 2}
\end{equation}
with
\begin{equation}
    \Delta\sigma^2 \coloneqq (\Delta\delta_a)^2 \sin^2\delta_a + (\Delta\alpha_a)^2 \cos^2\delta_a\,.
\end{equation}
We define the metric on the two-dimensional parameter space $(\alpha, \delta)$ as
\begin{equation}
    \intd\sigma^2 = e^{-2Y}(\intd X^2 + \intd Y^2)\,,
    \label{eq:metric on sky}
\end{equation}
with $X\coloneqq \alpha$ and $Y=-\log|\cos\delta|$\,. In this metric~\eqref{eq:metric on sky}, the contour of $\Delta\sigma^2$ becomes a circle, which allows us to easily place grid points.
In this work, we set the threshold at 
\begin{equation}
  \delta\Phi_\ast = 0.01\,.
\end{equation}
The condition~\eqref{eq:condition of grid points 2} depends on $f_\mathrm{gw}$. Here, we choose $f_\mathrm{gw}=100$ Hz, which gives an upper bound of the frequency band $f_\mathrm{up}$. When we analyze this frequency band lower than $f_\mathrm{up}$, we do not need a new set of grid points because the residual phase cannot be larger than the threshold determined with $f_\mathrm{gw} = f_\mathrm{up}$. With these choices, we obtain
\begin{equation}
    \Sub{N}{grid} = 5609178\,,
\end{equation}
grid points to cover the entire sky.

The value of $\delta\Phi_\ast$ is arbitrarily chosen. If $\delta\Phi_\ast$ increases, the number of grid points is reduced, meaning that the computational cost for the preprocessing would decrease. But, signal power may be lost because a larger $\delta\Phi_\ast$ allows a larger residual phase. If $\delta\Phi_\ast$ is set to a lower value, the signal will be more visible, but the computational cost of the preprocessing would increase. We will return to this point in Sec.~\ref{sec:computational cost}.

In the second step, the short-time Fourier transformation (SFT) is applied to the resampled strains to make spectrograms. Because a spectrogram is generated for each grid point, the number of spectrograms is $\Sub{N}{grid}$.
To avoid aliasing, each SFT segment is windowed by a Tukey window,
\begin{equation}
    w[m] = \begin{cases}
    \frac{1}{2} \left[ 1 - \cos\left( \frac{2\pi m}{\xi L} \right) \right]\,, & 0\leq m < \frac{\xi L}{2} \\
    1\,, &  \frac{\xi L}{2} \leq m \leq L-\frac{\xi L}{2} \\
    \frac{1}{2} \left[ 1- \cos\left( \frac{2\pi (L-m)}{\xi L}\right) \right]\,. & L-\frac{\xi L}{2} < m \leq L
    \end{cases}
\end{equation}
Here, $L$ is the window length and is given by
\begin{equation}
    L \coloneqq \Sub{T}{seg} \Sub{f}{s}\,,
\end{equation}
with the segment duration $\Sub{T}{seg}$.
$\xi$ is the parameter characterizing the steepness of the window edge, which we set to $\xi = 0.125$. A pixel value of a spectrogram is written as
\begin{equation}
    \tilde{s}_{ak}[j] = \frac{1}{L} \sum_{m=0}^{L-1} w[m] s_a[jL + m] e^{-2\pi i m k /L}\,.
    \label{eq: SFT s_jk}
\end{equation}
Here, $s_a[m]$ is the discrete strain data defined by
\begin{equation}
    s_a[m] \coloneqq s_a(m\Delta\tau)\,,
\end{equation}
with the time resolution $\Delta\tau = f_\mathrm{s}^{-1}$.
We refer to the frequency corresponding to the $k$-th frequency bin as $f_k$, given by 
\begin{equation}
    f_k = k \Delta f\,,
\end{equation}
where
\begin{equation}
    \Delta f = \frac{1}{\Sub{T}{seg}}\,
    \label{eq: frequency resolution}
\end{equation}
is the frequency resolution of SFT.
The number of segment is denoted by $\Sub{N}{seg}$ and is given by
\begin{equation}
  \Sub{N}{seg} = \frac{\Sub{T}{dur}}{\Sub{T}{seg}}\,.
\end{equation}
An index $j$ specifies an SFT segment.
For a given grid point $\ve{n}_a$, a spectrogram~\eqref{eq: SFT s_jk} can be regarded as a set
\begin{equation}
  \left\{ \tilde{s}_{ak} \mid k=1,2,\cdots,\Sub{N}{bin}\right\}\,
\end{equation}
of time series vectors
\begin{equation}
  \tilde{s}_{ak} \coloneqq (\tilde{s}_{ak}[0],\tilde{s}_{ak}[1],\cdots,\tilde{s}_{ak}[\Sub{N}{seg}-1])\,.
\end{equation}
Here, $\Sub{N}{bin}$ is the number of frequency bins we analyze and is given by
\begin{equation}
    \Sub{N}{bin} = \frac{\Sub{f}{up}}{\Delta f}\,.
\end{equation}
Each vector $\tilde{s}_{ak}$ has an index corresponding to a frequency bin. If the time resampling procedure perfectly demodulates the gravitational-wave signal, the signal power is contained in one frequency bin that corresponds to the source frequency. Hence, we analyze the data in each frequency bin and each grid point $\ve{n}_a$ separately.

\begin{figure}[t]
    \centering
    \includegraphics[width=8cm]{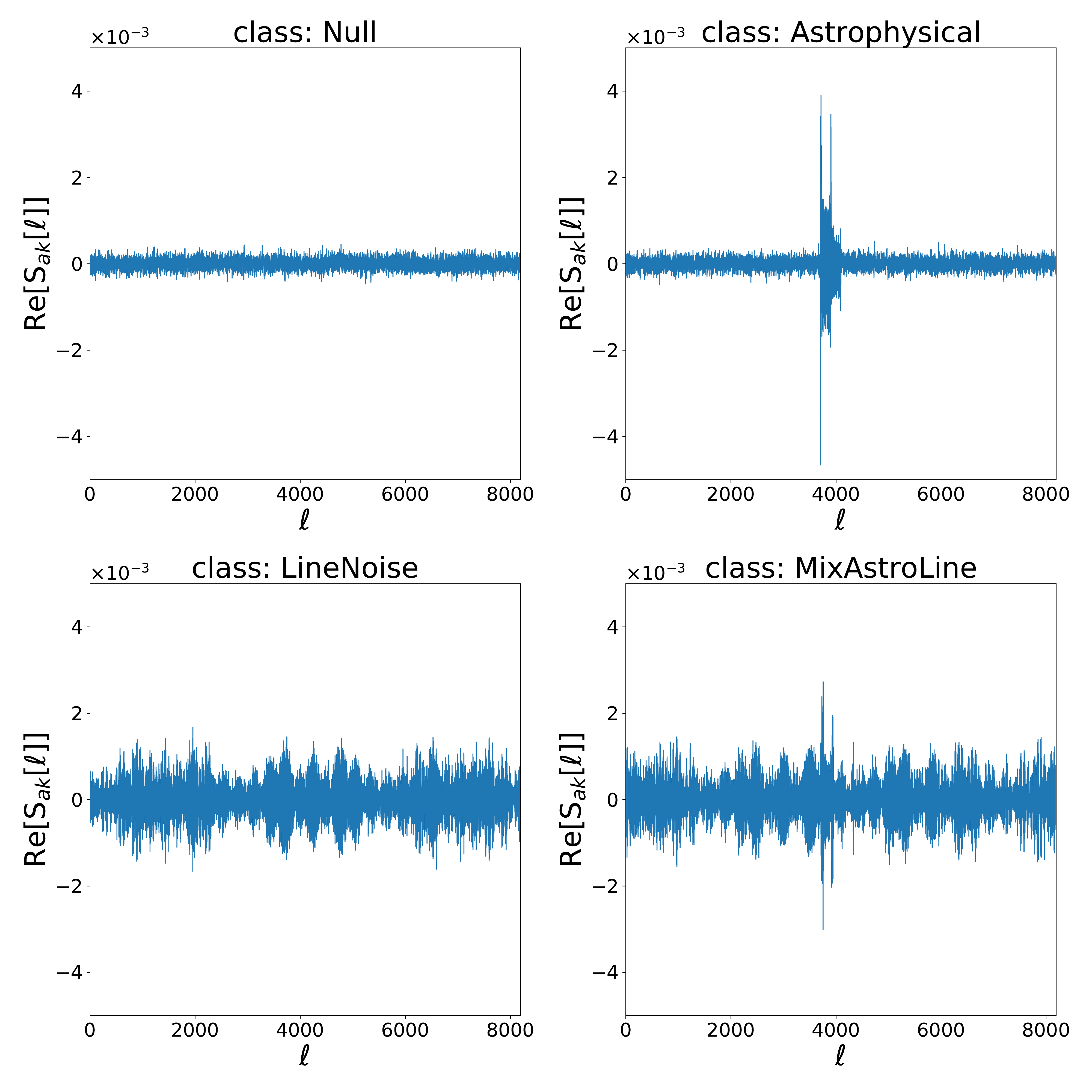}
    \caption{Examples of the real part of the doubly Fourier transformed signals $\mathsf{S}_{ak}$. From top left in the clockwise direction: only Gaussian noise, an astrophysical signal contaminated by Gaussian noise, an astrophysical signal contaminated by Gaussian noise and line noise, line noise and Gaussian noise. The amplitude of the astrophysical signal is set to $\log_{10} \hat{h}_0 = -1.0$ and the line noise is $\log_{10} \Super{\hat{h}}{line}_0 = 0.0$ (the amplitude parameters $\hat{h}_0$ and $\Super{\hat{h}}{line}_0$ are defined in Eq.~\eqref{eq: normalized amplitude} and~\eqref{eq: normalized amplitude of line noise}, respectively). Note that these values are an optimistic case. After the second Fourier transform, we shift $\ell$ as shown in Eq.~\eqref{eq: freqshift}. Therefore, if the GW frequency is close to the frequency of the SFT bin, the GW signal forms an excess around the center.}
    \label{fig: example of data}
\end{figure}

In the final step, another Fourier transform is applied to each time series vector. It is denoted by
\begin{equation}
  \mathsf{S}_{ak}[\ell] \coloneqq \frac{1}{\Sub{N}{seg}} \sum_{j=0}^{\Sub{N}{seg}-1} \tilde{s}_{ak}[j] e^{-2\pi ij\ell / \Sub{N}{seg}}\,.
  \label{eq: ell domain signal}
\end{equation}
In this expression, $\mathsf{S}_{ak}[\ell]$ for $0\leq \ell \leq N_\mathrm{seg}/2-1$ expresses the positive frequency components while the components of $N_\mathrm{seg}/2 \leq \ell \leq N_\mathrm{seg} -1$ is filled by the negative frequency part. To align the vector component in ascending order of the frequency, we shift the components as
\begin{equation}
    \pmat{\mathsf{S}_{ak}[0] \\ \mathsf{S}_{ak}[1] \\ \vdots \\ \mathsf{S}_{ak}[N_\mathrm{seg}/2-1] \\ \mathsf{S}_{ak}[N_\mathrm{seg}/2] \\ \vdots \\ \mathsf{S}_{ak}[N_\mathrm{seg}-1]}
    \to
    \pmat{\mathsf{S}_{ak}[N_\mathrm{seg}/2] \\ \mathsf{S}_{ak}[N_\mathrm{seg}/2+1] \\ \vdots \\ \mathsf{S}_{ak}[N_\mathrm{seg}-1] \\ \mathsf{S}_{ak}[0] \\ \vdots \\ \mathsf{S}_{ak}[N_\mathrm{seg}/2-1]}\,.
    \label{eq: freqshift}
\end{equation}
For simplicity, we interpret the elements of a vector
\begin{equation}
  \mathsf{S}_{ak} \coloneqq (\mathsf{S}_{ak}[0], \mathsf{S}_{ak}[1], \cdots, \mathsf{S}_{ak}[\Sub{N}{seg}-1])\,,
\end{equation}
as already ordered by the transformation~\eqref{eq: freqshift}.
Finally, we get a set
\begin{equation}
  \left\{ \mathsf{S}_{ak} \mid a = 1,2,\cdots,\Sub{N}{grid}; k=1,2,\cdots,\Sub{N}{bin} \right\}\,,
  \label{eq:set of vectors S}
\end{equation}
from a strain data.
Figure~\ref{fig: example of data} shows examples of preprocessed waveforms. The waveforms of Gaussian noise, astrophysical signals and line noise are significantly different due to each waveform's response to the time resampling procedure. This procedure accumulates  signal power in a small number of frequency bins; in contrast, it dilutes the power due to line noise in a wider frequency band. 

Now, we discuss the noise properties after the time resampling. We assume that the detector noise is stationary and Gaussian, and the time resampling does not change the statistical properties of the detector noise. The noise power spectrum $\Sub{S}{n}(f)$ of the detector noise is defined by
\begin{equation}
    \langle \tilde{n}(f) \tilde{n}^\ast(f') \rangle = \frac{1}{2} \Sub{S}{n}(f) \delta(f-f')\,,
\end{equation}
and we have defined the Fourier transform as
\begin{equation}
    \tilde{n}(f) = \int^\infty_{-\infty} \intd f\ n(t) e^{-2\pi i ft}\,.
    \label{eq: tilde n(f)}
\end{equation}
After whitening, the noises in different SFT segments are independent. Therefore, the noise power spectrum density is
\begin{equation}
    \langle \tilde{n}_{ak}[j]\tilde{n}^\ast_{ak}[j']  \rangle = \frac{1}{2\Sub{T}{seg}} \delta_{jj'}\,,
    \label{eq: <nn>}
\end{equation}
for any grid point $a$. Here, we denote the detector's Gaussian noise after the time resampling by
\begin{equation}
    n_a(\tau) \coloneqq n(t_a(\tau))\,,
\end{equation}
and its Fourier transform by
\begin{equation}
    \tilde{n}_{ak}[j] = \frac{1}{L} \sum_{m=0}^{L-1} n_a[jL + n] e^{-2\pi i m k /L}\,.
    \label{eq: SFT noise}
\end{equation}
We neglect the effect of the Tukey window. The Fourier transform in Eq.~\eqref{eq: SFT noise} differs from Eq.~\eqref{eq: tilde n(f)} by the normalization factor of $1/L$. Similarly to Eq.~\eqref{eq: ell domain signal}, we define
\begin{equation}
    \mathsf{N}_{ak}[\ell] \coloneqq \frac{1}{\Sub{N}{seg}} \sum_{j=0}^{\Sub{N}{seg}-1} \tilde{n}_{ak}[j] e^{-2\pi ij\ell / \Sub{N}{seg}}\,.
    \label{eq: noise in ell domain}
\end{equation}
The variance of $\mathsf{N}_{ak}[\ell]$ can be obtained as
\begin{align}
    &\quad \langle \mathsf{N}_{ak}[\ell] \mathsf{N}_{ak}^\ast[\ell'] \rangle \notag\\
    &= \frac{1}{\Sub{N}{seg}^2} \sum_{j,j'} \langle \tilde{n}_{ak}[j] \tilde{n}_{ak}^\ast[j'] \rangle e^{-2\pi i(j\ell - j'\ell')/\Sub{N}{seg}} \notag\\
    &= \frac{1}{2\Sub{N}{seg}\Sub{T}{seg}}\delta_{\ell\ell'}\,.
    \label{eq: noise variance in ell domain}
\end{align}
We generate simulated Gaussian noise in the transformed strain data by using Eq.~\eqref{eq: noise variance in ell domain}.\footnote{The Tukey window would reduce both the powers of signal and noise. Therefore, Eq.~\eqref{eq: noise variance in ell domain} overestimates the variance of noise. Because the CGW signal and the line noise are generated with considering the Tukey window, the SNR of simulated data could be underestimated. In this sense, our estimation of detection efficiency is conservative.}

\subsection{Line noise}
\label{sec: line noise}

Line noises are usually classified into three types: (1) perfectly sinusoidal line noise, (2) sinusoidal line noise with finite coherence time, and (3) comb line noise. Perfectly sinusoidal line noise is modeled by a sinusoidal function with a constant frequency. This is the simplest model of a line noise.

Perfectly sinusoidal line noise is modeled by
\begin{equation}
  \SupSub{n}{sin}{line}(t) = n_0 \cos(2\pi \Sub{f}{line} t  + \phi_0)\,,
  \label{eq: sinusoidal line noise}
\end{equation}
where $n_0$ is the line amplitude, $\Sub{f}{line}$ is the line noise frequency, and $\phi_0$ is the initial phase. The model~\eqref{eq: sinusoidal line noise} does not have a frequency, nor amplitude, modulation. The spectral density has a infinitely narrow peak at the frequency $\Sub{f}{line}$.

In practice, the frequency of line noise can change on a certain time scale. If line noise has a finite coherence time, the power of the line noise dissipates in a wide range of frequency bins. It leads to the suppression of line noise power contained in a preprocessed vector $\mathsf{S}_{ak}$. In this work, though, for simplicity we do not account for line noise with a finite coherence time.

Some instrumental disturbances could also cause multiple line noises with different frequencies that are evenly spaced, i.e., \textit{combs}. However, for most combs observed in the first and second observing runs of Advanced LIGO and Advanced Virgo \cite{LSC:2018vzm}, the spacing in frequency is much larger than the Doppler modulation. Therefore, each of the comb's teeth would be contained in a different frequency bin and would safely be regarded as a single line, so, we can focus here on perfectly sinusoidal line noise.

Let us remark on the amplitude of line noise we consider in the presented work. As we stated above, lines are assumed to be stable and monochromatic. They can be removed by whitening if their amplitudes are much larger than the Gaussian noise level. In reality, though, more sophisticated methods are required to remove lines because they have finite coherent times. Even stable lines cannot be completely removed if their amplitudes are comparable to the Gaussian noise level. Therefore, in this work, we assume the line noise amplitude to be in the range
\begin{equation}
    1.0 \leq n_0 \left( \frac{S_\mathrm{n}(f_k)}{1\mathrm{Hz}^{-1}} \right)^{-1/2} \leq 10.0\,.
\end{equation}

\section{Method}
\label{sec: method}

We use deep learning to discriminate the presence or absence of a GW signal and/or a sinusoidal line in Gaussian noise. The fundamentals of deep learning are summarized in Appendix.~\ref{sec:neural network}. In deep learning, we can use a neural network to extract data features and give a prediction for newly-obtained data. Here, we construct a convolutional neural network (CNN) to classify the input vectors $\mathsf{S}_{ak}$ in four classes: (1) only Gaussian noise (\verb|Null|), (2) astrophysical signal injected into Gaussian noise (\verb|Astrophysical|), (3) sinusoidal line noise injected into Gaussian noise (\verb|LineNoise|), and (4) astrophysical signal in the presence of both sinusoidal line noise and Gaussian noise (\verb|MixAstroLine|). Our CNN is trained to predict the probabilities~\eqref{eq: softmax layer} that certain strain data fall into each class.

Using the probabilities that the CNN has predicted, we then need to decide on a definition of ``detection'' of an astrophysical signal. Here, we choose to The common choice is that the data are classified into the class for which the CNN gives the largest probability. We assume that a vector contains an astrophysical signal if the CNN classifies the vector as \texttt{Astrophysical} or \texttt{MixAstroLine}, while we try another definition of ``detection'' later.

We use the term ``candidates'' to indicate a set of vectors determined to have an astrophysical signal based on the procedure described above. Each candidate is characterized by an SFT frequency bin and a grid point, as well as the vector $\mathsf{S}_{ak}$, whose indices describe the frequency bin $k$ and the grid point $a$.

\subsection{CNN architecture}
\begin{table}[h]
    \caption{\label{tab: structure of cnn}
    Structure of the CNN we used in this work. The first column shows types of layers. Here, we separately list the activation functions. Roughly speaking, the CNN can be divided into two blocks. The first block consists of convolutional layers, pooling layers, and activation functions. The second block comprises the fully-connected layers, activation functions, and a softmax layer. Before the first fully-connected layers, the transformation called \textit{flattening} is applied. It transforms a two-dimensional tensor into a one-dimensional vector. The second column shows the output size of the layer. For the layers before the flattening, the output is a two-dimensional tensor and its shape is described by two numbers. First number shows the number of channels while the second number is the length of data. The third column gives the kernel sizes of the convolutional layers and the pooling layers, while the last layer shows the number of tunable parameters. The first row shows the input vector that has the length of 8192 and two channels. The CNN has six convolutional layers. The number of tunable parameters is calculated by Eq.~\eqref{eq:tunable params fully connected layer} and~\eqref{eq: tunable parmas conv}. The total number of tunable parameters is 4171170.}
    \begin{ruledtabular}
    \begin{tabular}{llll}
        Layer & Output size & kernel size & \# of parameters \\ \hline
        (Input) & (2, 8192) & - & - \\
        1D convolutional & (16, 8177) & 16 & 528 \\
        ReLU & (16, 8177) & - \\
        1D convolutional & (16, 8162) & 16 & 4112 \\
        ReLU & (16, 8162) & - & - \\
        Max pooling & (16, 2040) & 4 & - \\
        1D convolutional & (32, 2033) & 16 & 4128 \\
        ReLU & (32, 2033) & - & - \\
        1D convolutional & (32, 2026) & 16 & 8224 \\
        ReLU & (32, 2026) & - & - \\
        Max pooling & (32, 506) & 4 & - \\
        1D convolutional & (64, 503) & 4 & 8256 \\
        ReLU & (64, 503) & - & - \\
        1D convolutional & (64, 500) & 4 & 16448 \\
        ReLU & (64, 500) & - & - \\
        Max pooling & (64, 125) & 4 & - \\
        Flattening & (8000,) & - & - \\
        Fully-connected & (512,) & - & 4096512 \\
        ReLU & (512,) & - & - \\
        Fully-connected & (64,) & - & 32832 \\
        ReLU & (64,) & - & - \\
        Fully-connected & (4,) & - & 260 \\
        Softmax & (4,) & - & -
    \end{tabular}
    \end{ruledtabular}
\end{table}

Table.~\ref{tab: structure of cnn} shows the structure of the CNN we used. It consists of six convolutional layers, three max-pooling layers, and three fully-connected layers. In the table, a ReLU transformation is counted as a layer, and a linear transform in the fully-connected layer is separated from the activation. 

The CNN takes the real part and the imaginary part of a vector $\mathsf{S}_{ak}$ as an input. Respecting Eq.~\eqref{eq: convolutional layer}, we write the input vector as
\begin{equation}
    x_{1j} = \RE{\mathsf{S}_{ak}[j]}\,,\qquad x_{2j} = \IM{\mathsf{S}_{ak}[j]}\,.
\end{equation}
It is known that normalizing input data accelerates and stabilizes the training~\cite{Goodfellow-et-al-2016}.
We normalize the input vector so that it has a mean of 0 and a standard deviation of unity:
\begin{equation}
    \hat{x}_{aj} = \frac{x_{aj} - \mu}{\sigma}\,,
    \label{eq: normalized input}
\end{equation}
where the mean is given by
\begin{equation}
    \mu \coloneqq \frac{1}{2N_\mathrm{in}} \sum_{a=1,2} \sum_{j=1}^{N_\mathrm{in}} x_{aj}\,,
\end{equation}
and the standard deviation is
\begin{equation}
    \sigma \coloneqq \sqrt{\frac{1}{2N_\mathrm{in}} \sum_{a=1,2} \sum_{j=1}^{N_\mathrm{in}} (x_{aj} - \mu)^2}\,.
\end{equation}

We employ the cross-entropy loss function~\eqref{eq: cross entropy loss}, use the Adam optimzer~\cite{Kingma:2014vow}, and implement the CNNs within the deep learning library \texttt{PyTorch}~\cite{Paszke2019}. The training and evaluation are carried out with a single GPU GeForce GTX1080Ti.
We trained the CNN for 300 epochs, and do not observe overfitting. Therefore, we use the CNN state at the end of the training.

\subsection{Data preparation}
In our work, we generate datasets in a limited frequency band, and the results (e.g., sensitivity, false alarm rate) are extrapolated lower frequencies. We use the frequency band of
\begin{equation}
    f_k - \frac{1}{2}\Delta f_\mathrm{gw} \leq f_\mathrm{gw} \leq f_k + \frac{1}{2} \Delta f\,,
    \label{eq: frequency range}
\end{equation}
with
\begin{equation}
    f_k = 100 \mathrm{\ Hz}\,.
\end{equation}
Equation~\eqref{eq: frequency range} is the width of the $k$-th frequency bin corresponding to 100 Hz.

In this work, we focus only on the selected frequency bin (i.e., $f_k=100$ Hz) and train on simulated monochromatic signals, 
\begin{equation}
    \dot{f} = 0\,.
\end{equation}

The sensitivity of the method is quantified by determining the minimum amplitude that an injected signal could be detected with a given detection probability and a specified false alarm probability. If we simply use $h_0$ as an indicator of the method's sensitivity, the sensitivity is affected by the detector noise level. To avoid such an effect, we normalize the amplitude as
\begin{equation}
    \hat{h}_0 \coloneqq h_0 \left( \frac{\Sub{S}{n}(f_k)}{1\mathrm{Hz}^{-1}} \right)^{-1/2}\,,
    \label{eq: normalized amplitude}
\end{equation}
where $\Sub{S}{n}(f_k)$ is the power spectral density of the detector's Gaussian noise at the reference frequency $f_k$. We use the normalized amplitude~\eqref{eq: normalized amplitude} when we generate the dataset and quantify our CNN's sensitivity.

\begin{table}[t]
    \centering
    \caption{\label{tab: source parameters}
    Source parameters. The range of normalized amplitudes is chosen so that it covers (1) the minimum limit of the detectable signal, and (2) sufficiently large signals so that the CNN can learn the signals efficiently.}
    \begin{ruledtabular}
    \begin{tabular}{ll}
        Description & Distribution \\ \hline
        Normalized amplitude $\hat{h}_0$ & Log uniform on $[10^{-2}, 10^1]$ \\
        Frequency $f_\mathrm{gw}$ & Uniform on $\left[ f_k - \frac{\Delta f}{2}, f_k + \frac{\Delta f}{2} \right]$ \\
        Inclination angle $\iota$ & Uniform on $[0,\pi]$ \\
        Right ascension $\alpha$ & Uniformly distributed on the sky \\
        Declination angle $\delta$ & Uniformly distributed on the sky \\
        Polarization angle $\psi$ & Uniform on $[0,2\pi]$ \\
        Initial phase $\phi_0$ & Uniform on $[0,2\pi]$ 
    \end{tabular}
    \end{ruledtabular}
\end{table}

Under this assumption, the \gwh signal can be characterized by seven parameters, i.e., a normalized amplitude $\hat{h}_0$, a frequency $f_\mathrm{gw}$, an inclination angle $\iota$, a right ascension $\alpha$, a declination angle $\delta$, a polarization angle $\psi$, and an initial phase $\phi_0$. Table~\ref{tab: source parameters} shows the distributions of source parameters that we sampled from to generate the training and the validation datasets. For the normalized amplitude, the upper limit of the range is set to be slightly larger than we expected in realistic situations. The neural network can learn the features of an astrophysical signal from data that contain signals with large amplitudes and gradually becomes able to capture the signature of lower amplitude signals. 
As for the source locations, we uniformly distribute the position on the sky. Here, we assume that the GW signal is significantly suppressed if the Doppler correction is not appropriate. Such a situation occurs when our chosen grid points lie far away from the source location. Therefore, when we generate the data of the \texttt{Astrophysical} and \texttt{MixAstroLine} classes, we pick the closest grid point to the source location.

\begin{table}[t]
    \centering
    \caption{\label{tab: line noise parameters}
    Line noise parameters.}
    \begin{ruledtabular}
    \begin{tabular}{ll}
        Description & Distribution \\ \hline
        Normalized amplitude $\Super{\hat{h}}{line}_0$ & Log uniform on $[1, 10]$ \\
        Frequency $f_\mathrm{line}$ & Uniform on $\left[ f_k - \frac{\Delta f}{2}, f_k + \frac{\Delta f}{2} \right]$ \\
        Initial phase $\phi_0$ & Uniform on $[0,2\pi]$ 
    \end{tabular}
    \end{ruledtabular}
\end{table}

We use Eq.~\eqref{eq: sinusoidal line noise} as the line noise model. It is characterized by an amplitude $n_0$, a frequency $\Sub{f}{line}$, and an initial phase $\phi_0$. The frequency $\Sub{f}{line}$ is also limited to the range of Eq.~\eqref{eq: frequency range}. The normalized amplitude~\eqref{eq: normalized amplitude} are employed instead of $n_0$ i.e.,
\begin{equation}
    \Super{\hat{h}}{line}_0 \coloneqq n_0 \left( \frac{\Sub{S}{n}(f_k)}{1\mathrm{Hz}^{-1}} \right)^{-1/2}\,.
    \label{eq: normalized amplitude of line noise}
\end{equation}
Table~\ref{tab: line noise parameters} shows the parameters characterizing line noise and how they are sampled. After generating sinusoidal line noise, we transform it using the time resampling procedure with the randomly chosen grid points.

We prepare 20000 GW signals, 20000 sinusoidal lines, and 20000 pairs of GW signals and lines for training. They correspond to \texttt{Astrophysical}, \texttt{LineNoise}, and \texttt{MixAstroLine} classes, respectively. In generating GW signals, we use the fact that the extrinsic parameters (amplitude, polarization angle, inclination angle, and initial phase) can be factored out of the CGW waveform. We therefore generate the waveform that depends only on $f_\mathrm{gw}$ and source position. For each iteration of training, we sample extrinsic parameters, include the effects of extrinsic parameters to determine the CGW waveform. 

A similar factorization can be done for the line noise waveform. We can factor out the amplitude and multiply the waveform by a randomly selected one in each iteration. This means that the line noise waveform depends only on $f_\mathrm{line}$. Before feeding the waveforms into the CNN, we inject them into Gaussian noise with the variance given by Eq.~\eqref{eq: noise variance in ell domain}.

For the \texttt{Null} class, we only give Gaussian noise to the CNN. Thus, we do not need to generate any data for \texttt{Null} class in advance. The validation data are generated by the same procedure, but the number of data is decreased to 2000 for each class.

\section{Results}
\label{sec: results}

\begin{figure}[t]
    \centering
    \includegraphics[width=8.5cm]{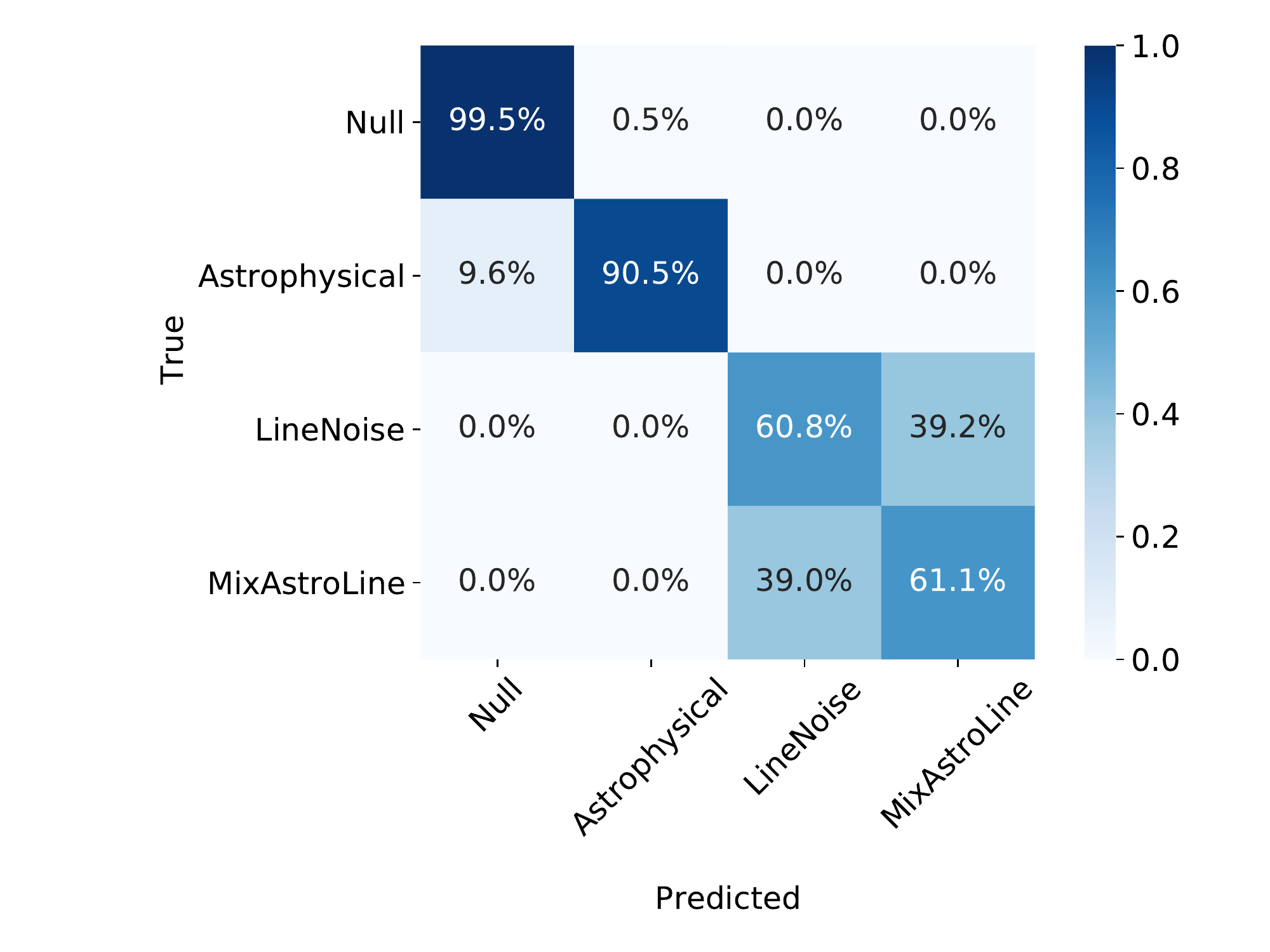}
    \caption{Confusion matrix of the CNN. This matrix quantifies the fraction of testing data that were classified correctly (diagonal elements) and incorrectly (off-diagonal elements).}
    \label{fig:confusion matrix}
\end{figure}

Figure~\ref{fig:confusion matrix} shows the confusion matrix of the trained CNN. We use 2000 test data for each class. The amplitude of gravitational wave is uniformly sampled from $\log_{10} \hat{h}_0 \in [-2.0, -1.0]$, and that of line noise is sampled from $\log_{10} \Super{\hat{h}}{line}_0 \in [0.0, 1.0]$. Here, we changed the range of the signal amplitude from that employed in the training data because we want to test our CNN for data with realistic amplitudes. Most of the data in the \verb|Null| class is correctly classified as the \verb|Null| class. The significant point of Fig.~\ref{fig:confusion matrix} is that the CNN can discriminate between the presence and the absence of line noises. The test events of the \verb|Null| class and the \verb|Astrophysical| class are not misclassified as the \verb|LineNoise| class or the \verb|MixAstroLine| class. On the contrary, the test data containing line noise are classified in the \verb|LineNoise| class or the \verb|MixAstroLine| class, and there are small confusions with the \verb|Null| class. From these results, it can be concluded that the CNN can tell apart a line from its absence.

\begin{table}[t]
    \caption{\label{tab:test null}
    Result for the case where only Gaussian noise exists. We use 20000 test events that contains only simulated Gaussian noise. Our CNN can classifies the Gaussian noise data with the accuracy of 99.34\%. The false alarm probability is 0.66\%.}
    \begin{ruledtabular}
    \begin{tabular}{lll}
        Predicted class & \# of events & Fraction [\%] \\ \hline
        Null & 19868 & 99.34\\
        Astrophysical & 132 & 0.66\\
        LineNoise & 0 & 0.0 \\
        MixAstroLine & 0 & 0.0
    \end{tabular}
    \end{ruledtabular}
\end{table}
From the top row of Fig.~\ref{fig:confusion matrix}, it is found that only 0.5\% of events that contain just Gaussian noise are classified as the \verb|Astrophysical| class. Furthermore, we use 20000 simulated Gaussian noise data for evaluating the false alarm probability for the Gaussian noise, as shown in Table.~\ref{tab:test null}. Among 20000 test noise data, 132 events are classified as the \verb|Astrophysical| class. The estimated false alarm probability to misclassify Gaussian noise as an astrophysical signal is 0.66\%, which is comparable to that estimated by 2000 test events. Even for 20000 events, we find no confusion between the line noise class and the mixed class.

\begin{figure}[t]
    \centering
    \includegraphics[width=8cm]{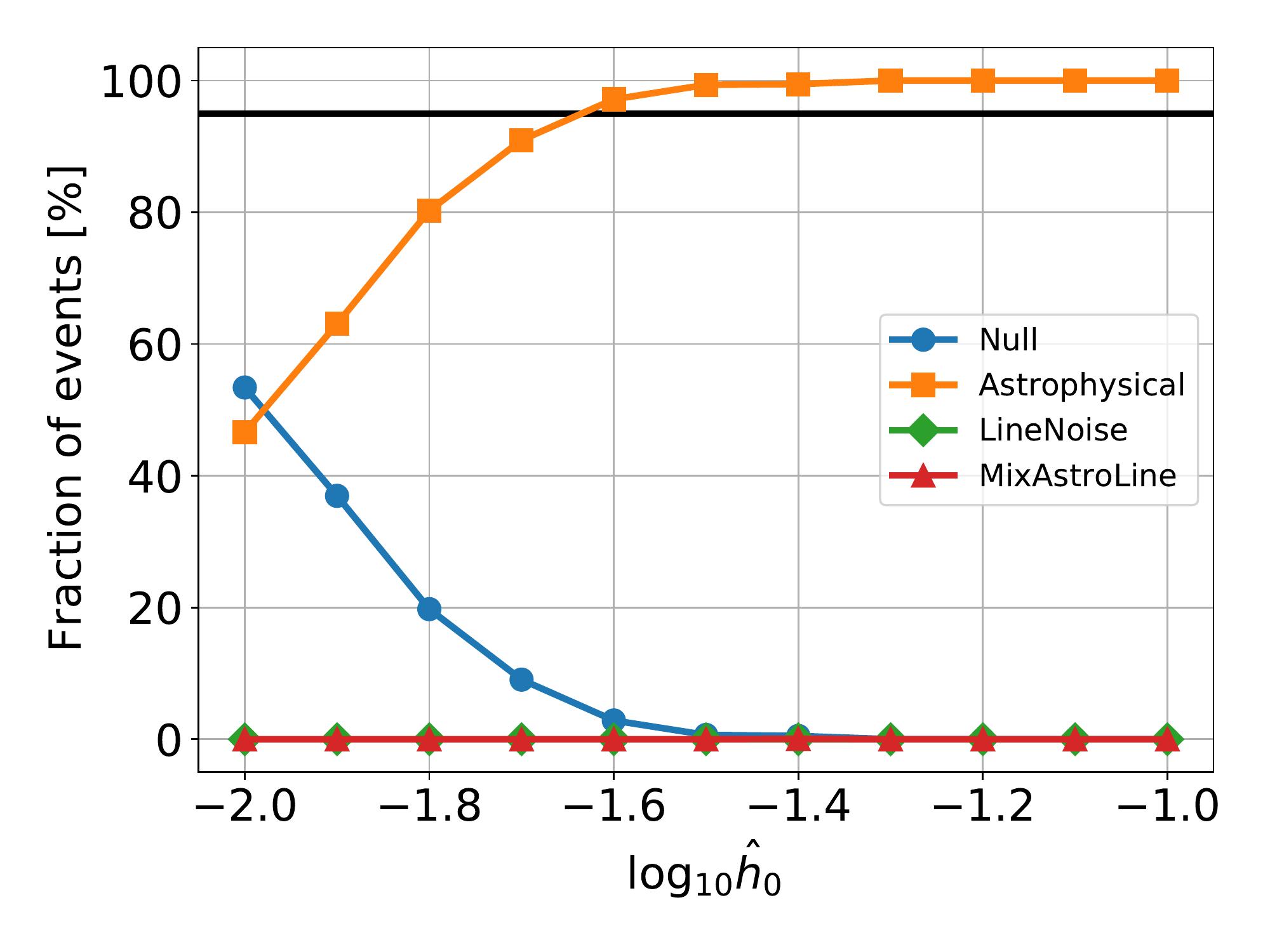}
    \caption{Detection efficiency of the CNN for astrophysical signals injected into Gaussian noise. The horizontal axis shows the logarithm of the amplitude, and the vertical axis is the fraction of events. Orange squares indicate the detection probability of astrophysical signals. For $\log_{10} \hat{h}_0 \gtrsim -1.64$, the detection probability exceeds 95\%. The detection probability decreases as the amplitude decreases. The results of the \texttt{LineNoise} class (green diamonds) and the \texttt{MixAstroLine} class (red triangles) are overlapped.}
    \label{fig:accuracy astro}
\end{figure}
We study the detailed result for the \verb|Astrophysical| class. With varying the normalized amplitude $\log_{10}\hat{h}_0$ from -2.0 to -1.0 with the step of 0.1, we prepare eleven datasets corresponding to the respective values of the amplitude. Each data set consists of 2000 injections. We apply the trained CNN to each dataset and count the number of detected events for each predicted class. Figure~\ref{fig:accuracy astro} shows the fraction of events as a function of the normalized amplitude. The detection probability exceeds 95\% for $\log_{10}\hat{h}_0 \gtrsim -1.64$. We also quote our results in terms of the so-called \textit{sensitivity depth}, which is defined as
\begin{equation}
    \mathcal{D} \coloneqq \frac{ \sqrt{ \Sub{S}{n}(f_k) / 1\mathrm{Hz^{-1}} } }{h_0} = (\hat{h}_0)^{-1}\,.
\end{equation}
In terms of the sensitivity depth, our CNN has a sensitivity of 
\begin{equation}
    \mathcal{D}^\mathrm{95\%} \simeq 43.9\,.
    \label{eq: D95 without line}
\end{equation}
The LIGO/Virgo collaboration has carried out all-sky searches for isolated neutron stars using data from LIGO/Virgo's third observation data~\cite{LIGOScientific:2022pjk}, which results in upper limits on the gravitational-wave strain amplitude. We compare the sensitivity depths of the standard methods and our method in Table.~\ref{tab:sensitivity comparison}. It shows that our neural network can outperform the Time-domain $\mathcal{F}$-statistic and the SOAP. Furthermore, our method has comparable sensitivity to the FrequencyHough and the SkyHough. We emphasize, however, that they search over different parameter spaces: the standard method surveys a wide range of $\dot{f}$, while our method focuses on quasi-monochromatic waves. The duration of the signal is also different; O3 data has the duration of $\sim$ 11 months $\sim 2.9\times10^7$ sec, and our method assumes that signals last for $2^{24} \sim 1.6\times10^{7}$ sec.
\begin{table}[t]
    \centering
    \caption{Comparison of the sensitivity depths of the standard all-sky search methods and our method. For FrequencyHough and Time-domain $\mathcal{F}$-statistic, the upper limits on the amplitude $h_0^\mathrm{95\%}$ are presented in \cite{LIGOScientific:2022pjk}. We converted them into $\mathcal{D}^\mathrm{95\%}$ assuming $\sqrt{\Sub{S}{n}(f)}=5.2\times10^{-24} \mathrm{[Hz^{-1}]}$ that is shown in Fig.~6 of~\cite{LIGOScientific:2022pjk}. For SkyHough and Time-domain $\mathcal{F}$-statistic, we read the values respectively from Fig.~11 and Fig.~13 of~\cite{LIGOScientific:2022pjk} that show their upper limit on the amplitude. We stress that the parameter region and the strain duration are different depending on the method.}
    \label{tab:sensitivity comparison}
    \begin{ruledtabular}
    \begin{tabular}{lll}
        Method & Frequency band & $\mathcal{D}^\mathrm{95\%}$ \\ \hline
        FrequencyHough & at 100 Hz & 42$\sim$ 43 \\
        SkyHough & at 116.5 Hz & 47.2\\
        Time-domain $\mathcal{F}$-statistic & at 100 Hz & 26$\sim$52\\
        SOAP & on 40$\sim$500 Hz & 9.9 \\
        Our method & $\lesssim$ 100 Hz & 43.9
    \end{tabular}
    \end{ruledtabular}
\end{table}

\begin{figure}[t]
    \centering
    \includegraphics[width=8cm]{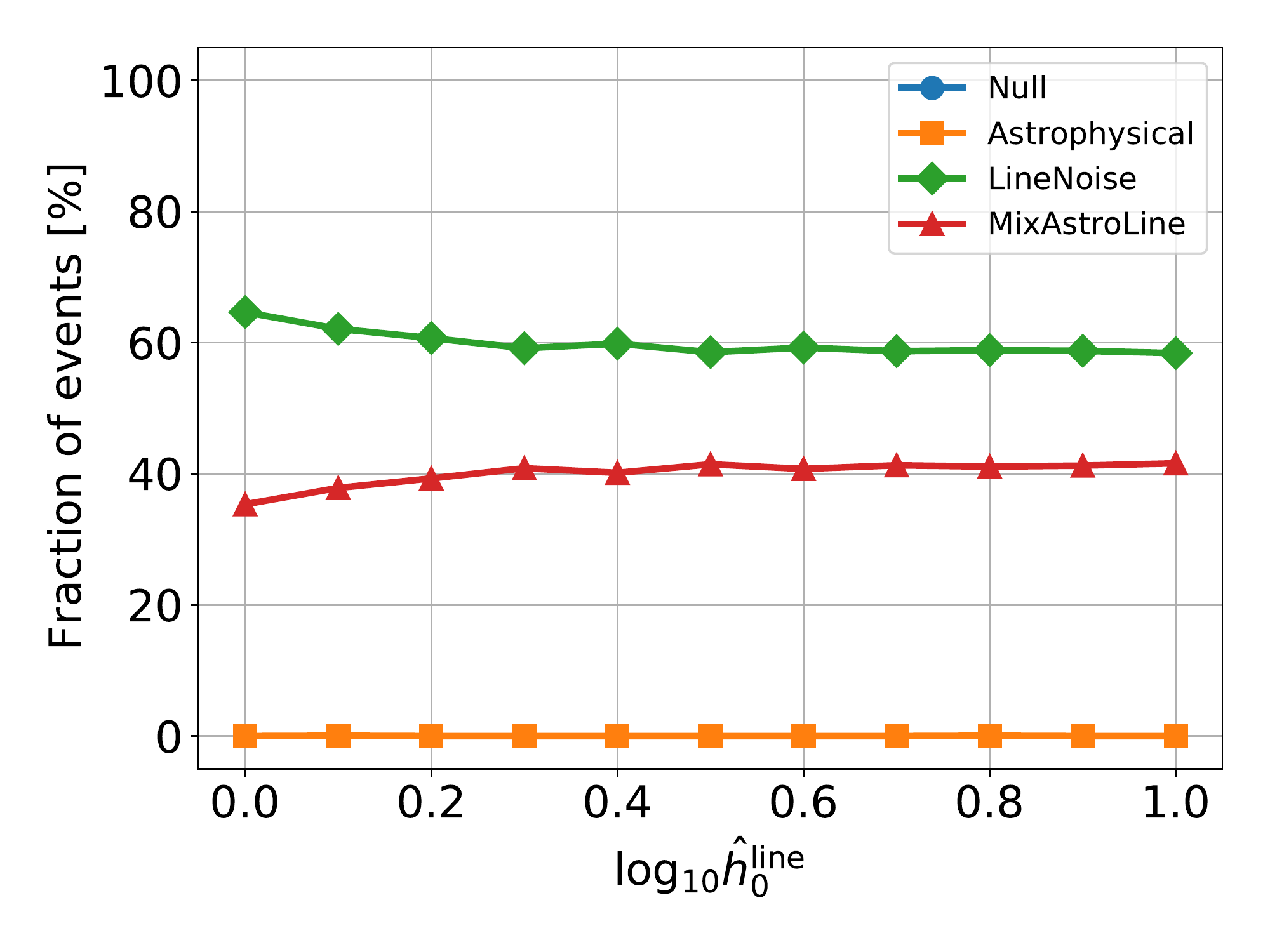}
    \caption{Classification results for test data containing only line noise with Gaussian noise. For any amplitude, the fraction of correctly classified events is about 60\% (green diamonds). The misclassification as the \texttt{MixAstroLine} class (red triangles) occurs for 40\% of test data. The number of misclassifications as the \texttt{Null} class (blue circles) and the \texttt{Astrophysical} class (orange squares) are almost zero. Their markers are overlapped.}
    \label{fig:results for line noise}
\end{figure}

Whereas the \verb|Astrophysical| class and the \verb|Null| class are classified correctly, the events contaminated by the line noise are not. The false alarm probability that the line noise data is classified in the \verb|MixAstroLine| class is estimated to be 39.2\%. To test the CNN for line noise data, we prepare eleven data sets corresponding to different amplitudes of the line noise. Each data set contains 2000 lines injected into Gaussian noise. Figure~\ref{fig:results for line noise} shows the classification result for the test data of the \verb|LineNoise| class as a function of the line noise amplitude $\log_{10} \Super{\hat{h}}{line}_0$. The classification results are almost constant for any value of $\log_{10} \Super{\hat{h}}{line}_0$. We can interpret this result as follows: we have injected  line noise events with amplitudes much larger than the Gaussian noise. Therefore, the overall amplitude of the line noise would disappear by normalization (see Eq.~\eqref{eq: normalized input}), with the result that the sensitivity of the CNN does not depend on the line noise amplitude, as shown in Fig.~\ref{fig:results for line noise}.

\begin{figure}[t]
    \centering
    \includegraphics[width=8.5cm]{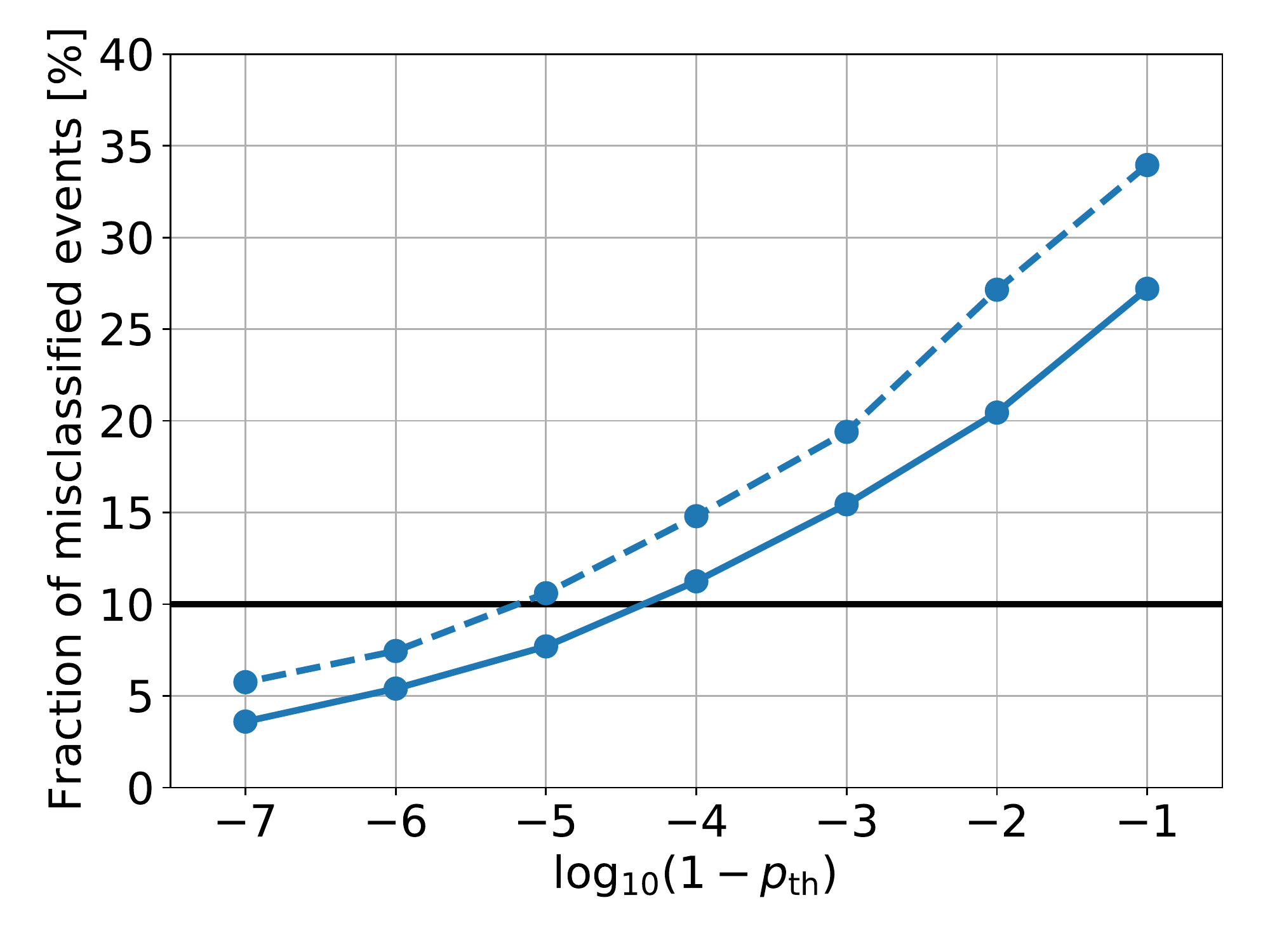}
    \caption{False alarm probabilities as a function of the threshold $\Sub{p}{th}$. The horizontal axis corresponds to $\Sub{p}{th}$. The vertical axis shows the fraction of \texttt{LineNoise} events which are misclassified as \texttt{MixAstroLine}. The dashed lines and the solid lines respectively present the cases of $\log_{10} \Super{\hat{h}}{line}_0 = 0.0$ and $1.0$. Black horizontal line corresponds to the misclassification probability of 10\%. If we set $\Sub{p}{th} = 1 - 10^{-6}$, we can suppress the false alarm probability less than 10\%.}
    \label{fig:false alarm line noise}
\end{figure}

While the CNN can discriminate the presence and the absence of a line, it cannot find the astrophysical signal when line noise contaminates. As shown in Fig.~\ref{fig:confusion matrix}, line noise is misclassified as the \verb|MixAstroLine| with the false alarm probability of $\sim$ 40\%. The false alarm probability could be suppressed by changing the detection criterion. As stated in the Sec.~\ref{sec: method}, we define a ``detection'' as when the predicted probability of \texttt{MixAstroLine} (or \texttt{Astrophysical}) class dominates others. Here, we introduce a new criterion, given by
\begin{equation}
    \Sub{p}{th} \leq \Sub{p}{Mix}\,,
\end{equation}
where $\Sub{p}{Mix}$ is the CNN predicted probability of the \verb|MixAstroLine| class. Figure~\ref{fig:false alarm line noise} shows the false alarm probabilities with various values of $\Sub{p}{th}$. In order to achieve a false alarm probability that is less than 10\% for data contaminated by line noise, we need to set $\Sub{p}{th} = 1-10^{-6}$. This detection threshold is used in the rest of the paper.

\begin{figure}[t]
    \centering
    \includegraphics[width=8cm]{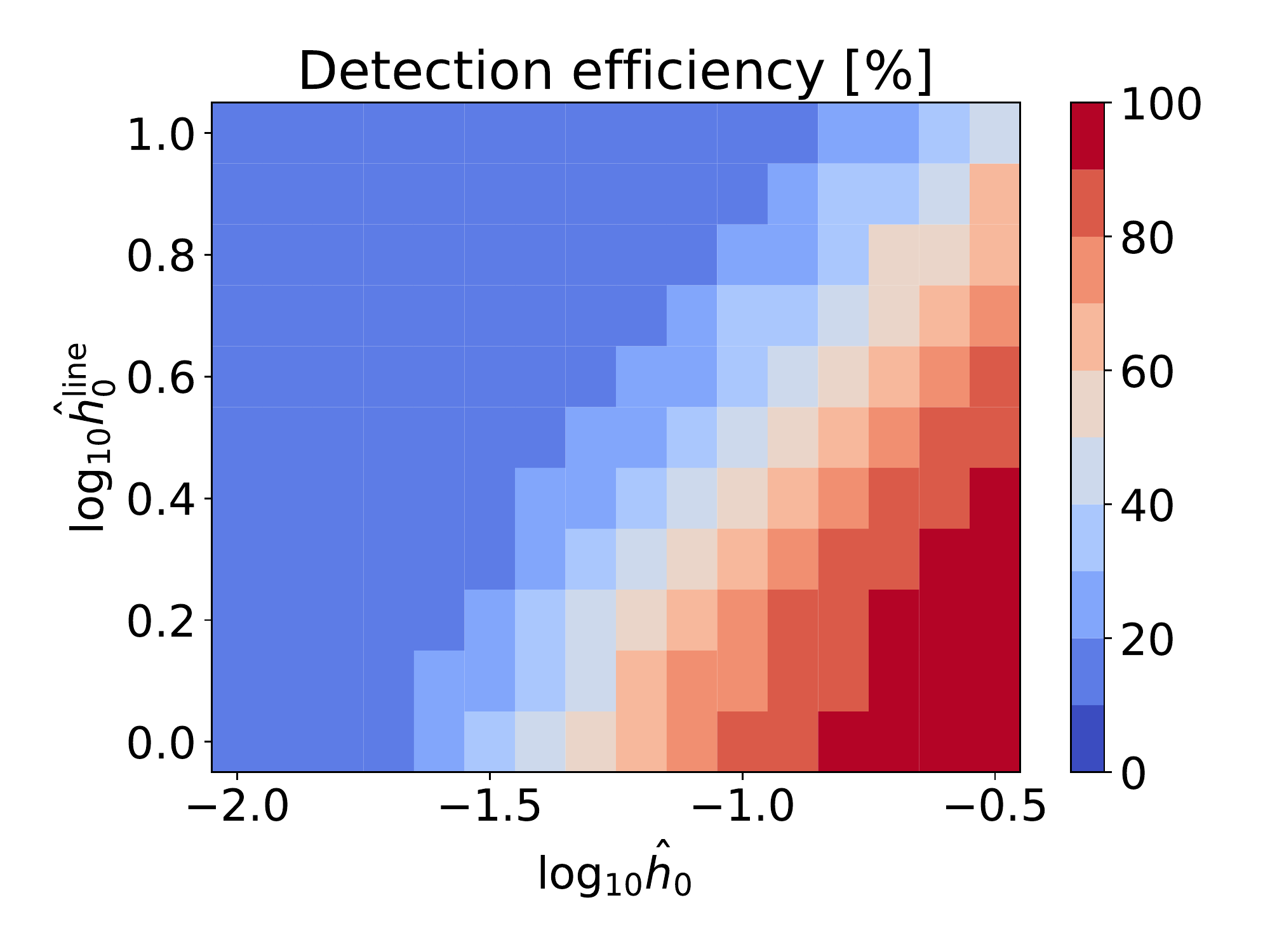}
    \caption{Detection probabilities of an astrophysical signal coexisting with line noise. The horizontal and vertical axes show the normalized amplitudes of astrophysical signals and line noise, respectively. In this figure, we set the threshold $\Sub{p}{th} = 1 - 10^{-6}$. In most regions, the detection probabilities are less than 50\%. The maximum detection probability is 96.1\% at $(\log_{10} \hat{h}_0, \log_{10} \Super{\hat{h}}{line}_0) = (-0.5, 0.0)$.}
    \label{fig:detection probability mix}
\end{figure}

Figure~\ref{fig:detection probability mix} shows the detection efficiency of the \texttt{MixAstroLine} signals Comparing to the case where the line noise is absent, the efficiency is degraded because of the line noise. We estimate the sensitivity depth
\begin{equation}
    \Super{\mathcal{D}}{95\%} \simeq 3.62
    \text{ for } \log_{10} \hat{h}^{\mathrm{line}}_0 = 0.0\,,
\end{equation}
which is only $\sim$ 8.2\% of that of the line noise is absence (see Eq~\eqref{eq: D95 without line}).

\begin{figure}[t]
    \centering
    \includegraphics[width=8.5cm]{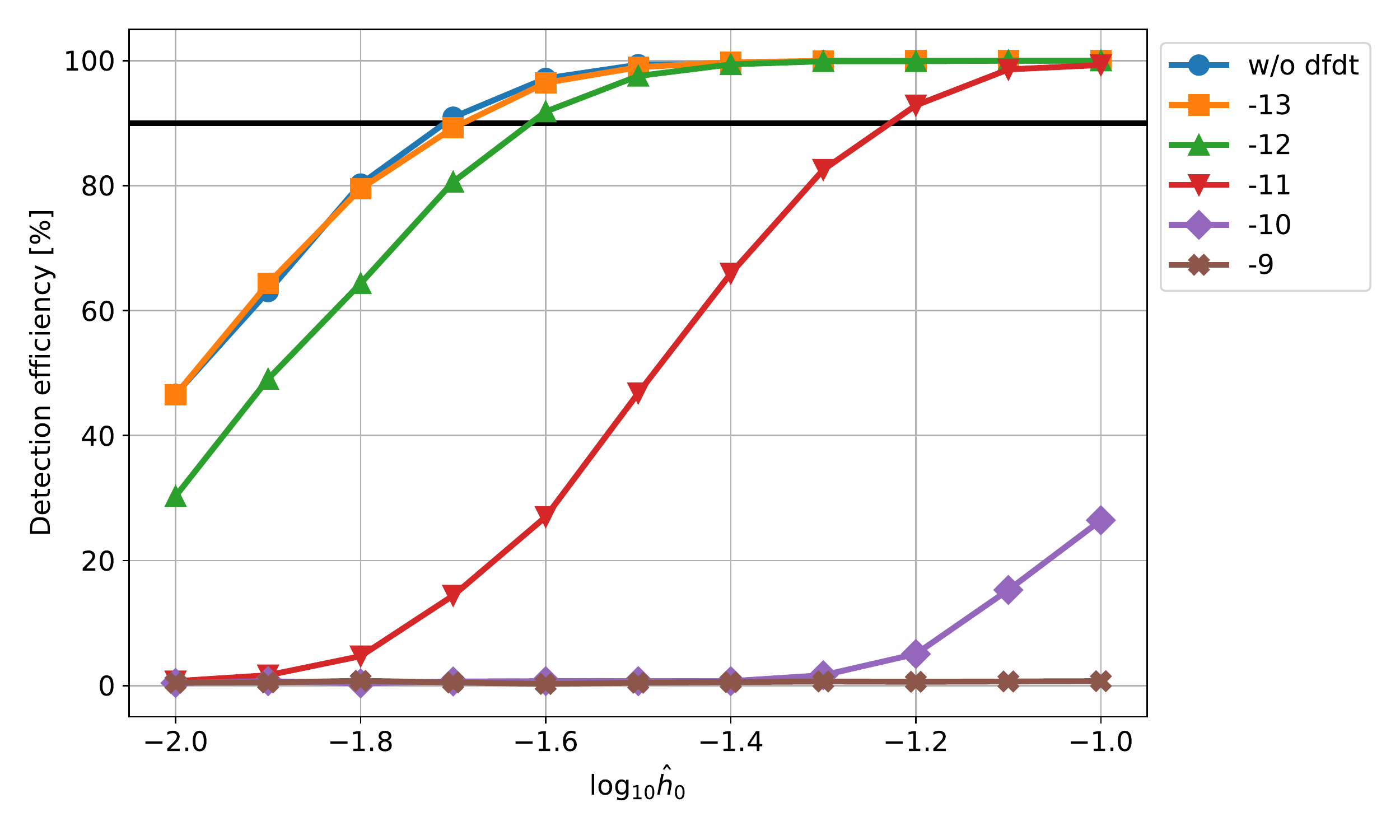}
    \caption{Detection probability of the signals with nonzero frequency derivatives. For $\dot{f}=-1.0\times 10^{-13}$ Hz/sec which are shown by orange squares, the sensitivity is not degraded compared with $\dot{f}=0$ case (blue circles). The detection probability starts to diminish from $\dot{f}=-1.0\times10^{-12}$ Hz/sec (green up triangles). For $|\dot{f}|\lesssim 10^{-11}$ Hz/sec, the sensitivity significantly reduced.}
    \label{fig:with dfdt}
\end{figure}

Realistic \gwh sources naturally have intrinsic frequency evolution as they are modeled in Eq.~\eqref{eq: signal phase}. Therefore, we test the 4-class CNN also for signals with non-zero $\dot{f}$. Different datsets are generated, each with the fixed $\dot{f}$: $\dot{f} = -10^{-13}, -10^{-12}, -10^{-11}, -10^{-10}$ and $-10^{-9} \mathrm{Hz/sec}$. For each $\dot{f}$, we prepare 2000 test data and evaluate the detection probability, and show the classification results in Fig.~\ref{fig:with dfdt}. For the data with $|\dot{f}|$ smaller than $10^{-12} \mathrm{Hz/sec}$, the CNN's performance is not much degraded. Especially for $\dot{f} = -10^{-13} \mathrm{Hz/sec}$, the detection probability is comparable to that of $\dot{f}=0$ case for all amplitudes. On the other hand, the performance becomes worse as the frequency derivative exceeds $|\dot{f}| = 10^{-11} \mathrm{Hz/sec}$. It can be understood as follows: as explained in Sec.~\ref{sec:waveform and preprocess}, the input data should contain the signal power with an SFT frequency bin. With nonzero $\dot{f}$, however, the frequency track might cross a number of frequency bins, spreading the signal power over multiple frequency bins. We, therefore, expect that signals with higher $\dot{f}$ cannot be detected as efficiently by the CNN as those with lower $\dot{f}$. 

Quantitatively, the frequency width of a bin is
\begin{equation}
    \Delta f = \frac{1}{\Sub{T}{seg}} \simeq 4.88 \times 10^{-4} \mathrm{Hz}\,.
\end{equation}
The frequency change from the initial time across $\Sub{T}{dur}$ can be estimated as
\begin{equation}
    \delta f \sim \Sub{T}{dur} \dot{f} \sim 10^{-4} \mathrm{Hz} \left( \frac{\dot{f}}{10^{-11} \mathrm{Hz/sec}} \right)\,.
\end{equation}
Roughly speaking, if $\delta f \lesssim \Delta f$, the signal power is still contained in one frequency bin. Thus, we expect that the CNN is applicable with comparable accuracy to that achieved in the $\dot{f}=0$ case. On the other hand, if $\Delta f \lesssim \delta f$, the signal power dissipates into several frequency bins. Thus, the CNN's performance degrades when $|\dot{f}| \gtrsim 10^{-11} \mathrm{Hz/sec}$.

\section{Computational cost}
\label{sec:computational cost}

In this section, we evaluate the computational cost of each processing step. First, we estimate the computational cost of the preprocess. The most expensive part of the preprocess is the SFT for making the spectrogram and the Fourier transform to obtain a set of vectors $\{\mathsf{S}_{ak}\}$. We assume that the cost of the time resampling is negligible compared to the SFT and the Fourier transform. For each grid point, we perform the SFT and the Fourier transform. The computational cost of taking SFTs can be estimated by
\begin{equation}
    \Sub{\mathcal{N}}{SFT} = \Sub{N}{seg} \cdot 5\Sub{f}{s} \Sub{T}{seg} \log_2 [\Sub{f}{s} \Sub{T}{seg}]\,.
    \label{eq: computational cost of SFT}
\end{equation}
Here, we evaluate the number of data points contained in an SFT segment as $\Sub{f}{s} \Sub{T}{seg}$. Using the values listed in Tab.~\ref{tab:preprocess parameters}, we estimate
\begin{equation}
    \Sub{\mathcal{N}}{SFT} \simeq 1.80 \times 10^{12}\,,
\end{equation}
in the unit of the number of floating point operations. Similarly, the computational cost of the Fourier transform for achieving a set of vectors $\{\mathsf{S}_{ak}\}$ can be estimated as
\begin{align}
    \Sub{\mathcal{N}}{Fourier} &= \Sub{N}{bin} \left(5\Sub{N}{seg} \log_2 \Sub{N}{seg}\right) \notag \\
    &\simeq 1.09 \times 10^{11}\,.
    \label{eq: computational cost of Fourier}
\end{align}
Combining Eqs.~\eqref{eq: computational cost of SFT} and~\eqref{eq: computational cost of Fourier}, we obtain the computational cost of the preprocess,
\begin{align}
    \Sub{\mathcal{N}}{preprocess} &= \Sub{N}{grid}(\Sub{\mathcal{N}}{SFT} + \Sub{\mathcal{N}}{Fourier}) \notag \\
    &\simeq 1.07 \times 10^{19}\,.
    \label{eq: computational cost of preprocess}
\end{align}

We evaluate the computational time of the CNN by extrapolating the measured value for a small subset consisting of the test data. With a single GPU (GTX1080Ti), we measure the computational time to process $10^5$ data five times. We obtained their averaged time of 8.8742 sec and the standard deviation of 0.0237 sec. The total number of the vectors $\{\mathsf{S}_{ak}\}$ to be processed is 
\begin{equation}
    \Sub{N}{vec} = \Sub{N}{grid} \cdot \Sub{N}{bin} = 1.15 \times 10^{12}\,.
\end{equation}
Therefore, the estimated time to process all vectors is
\begin{equation}
    \Sub{T}{CNN} \simeq 1.02 \times 10^8 \mathrm{\ [sec]}\,.
\end{equation}
Although this is longer than the total duration $\Sub{T}{dur}$ by an order of magnitude, we expect this can be suppressed to a negligible level by taking into account the development of hardware and the use of multiple GPUs in parallel. \footnote{Assuming the use of 10 GPUs with twice faster than GTX1080Ti that is used in this work, we have computational time $T_\mathrm{CNN} \simeq 5.0 \times 10^6 \mathrm{[sec]}$.}

In Table~\ref{tab:computational time comparison}, we compare the computational cost, in units of core-hours, of our method to that from the standard all-sky search pipelines employed in LIGO/Virgo's second observing run ~\cite{LIGOScientific:2019yhl} . As in~\cite{LIGOScientific:2019yhl}, we assume the hardware Intel E5-2670 that has a clock frequency of 2.6 GHz and carries out eight floating-point operations per clock. The computational speed is 20.8 GFlops per core. Using Eq.~\eqref{eq: computational cost of preprocess}, we estimate the computational time by
\begin{equation}
    \frac{\mathcal{N}_\mathrm{preprocess}}{20.8\ \mathrm{[GFlops]}} \simeq 1.4 \times 10^{5}\ \mathrm{[corehr]}\,.
\end{equation}
It shows that our method is computationally more efficient by one or two order of magnitude than the standard methods in which deep learning is not employed. Again, we stress that the parameter region and the duration of the strain data are different depending on the method.

\begin{table}[t]
    \centering
    \caption{Comparison of the computational time of the standard methods and our method. We estimate the core-hour with the spec of Intel E5-2670; the clock frequency is 2.6 GHz, eight operations per clock leading to the computational speed of 20.8 GFlops per core. This computational time includes only floating-point operations. We take these values from~\cite{LIGOScientific:2019yhl}, except that the computational time of SOAP is taken from~\cite{Bayley:2020zfa}. We do not consider input/output (I/O) time.}
    \label{tab:computational time comparison}
    \begin{ruledtabular}
    \begin{tabular}{ll}
        Method & core-hour \\ \hline
        FrequencyHough & $9 \times 10^{6}$ \\
        SkyHough & $2.5 \times 10^{6}$ \\
        Time-domain $\mathcal{F}$-statistic & $2.4 \times 10^{7}$ \\
        SOAP & $1-2 \times 10^{2}$\\
        Our method & $1.4 \times 10^{5}$ 
    \end{tabular}
    \end{ruledtabular}
\end{table}

\begin{figure}[t]
    \centering
    \includegraphics[width=8cm]{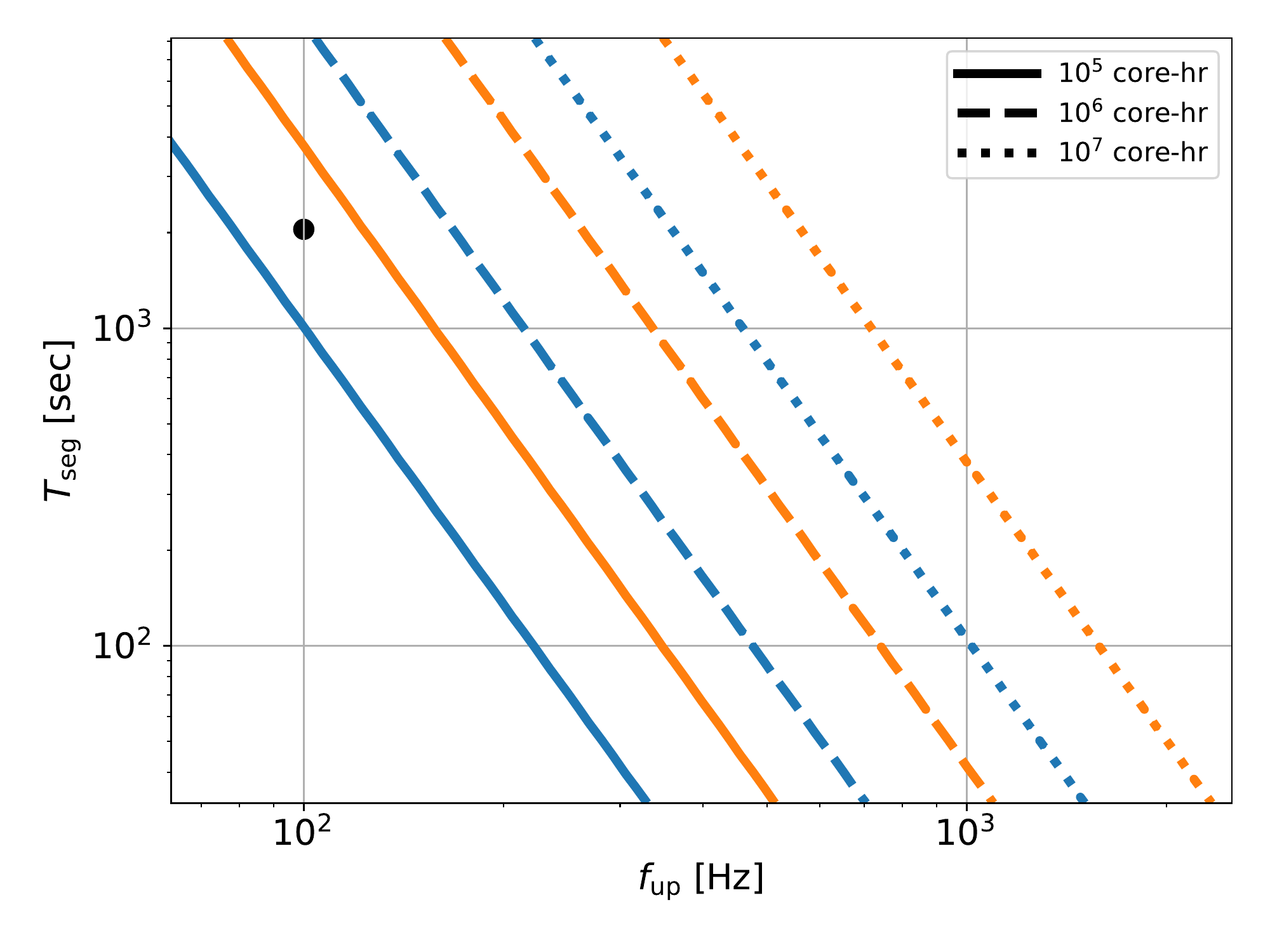}
    \caption{Computational cost of preprocessing with various parameter values. The horizontal axis is $\Sub{f}{up}$, and the vertical axis is $\Sub{T}{seg}$. Blue lines and orange lines show the case of $\delta\Phi_\ast = 0.01$ and 0.02, respectively. Solid, dashed and dotted lines respectively correspond to the contour lines of $10^5$, $10^6$, and $10^7$ core-hour. We assume a CPU Intel E5-2670 with a computational speed of 20.8 GFlops. The black dot shows our current choice of parameters $\Sub{f}{up} = 100$ Hz and $\Sub{T}{seg}=2048$ sec.}
    \label{fig:comp cost various params}
\end{figure}

Before ending this section, we mention how the computational cost of the preprocess depends on the various parameters governing our method. We focus on three parameters: the phase resolution $\delta\Phi_\ast$, the duration of the SFT segment $\Sub{T}{seg}$, and the upper bound of the frequency band we explore $\Sub{f}{up}$. Figure.~\ref{fig:comp cost various params} shows the computational cost of the preprocess with various values of $\delta\Phi_\ast$, $T_\mathrm{seg}$, and $f_\mathrm{up}$. To create this figure, we assume that the sampling frequency is set to $\Sub{f}{s} = 10 \Sub{f}{up}$. From this figure, we need to choose a low $\Sub{f}{up}$, a short $\Sub{T}{seg}$, and a high $\delta\Phi_\ast$ in order to reduce the computational cost of the preprocess. In the standard methods, $\Sub{f}{up}$ is usually set to $\sim 10^{3}$ Hz. We can make our method applicable for such frequency bands by choosing $\Sub{T}{seg}$ and $\delta\Phi_\ast$ appropriately. If we want to suppress the computational time to $10^7$ core-hour to explore the frequency band up to $10^3$ Hz, we should set $\Sub{T}{seg} \simeq 100$ sec or 400 sec for $\delta\Phi_\ast$ = 0.01 and 0.02, respectively. Changing the parameters affects not only the computational cost but also the performance of our CNN. It is important to study the dependence of the performance, but we leave it as future work.

\section{Conclusion}
\label{sec:conclusion}
CGWs from asymmetrically rotating neutron stars or depleting boson clouds around rotating black holes are exciting targets of the ground-based interferometers. However, there are two main difficulties for all-sky searches of CGWs: (1) the computational cost due to the Doppler effect and (2) the presence of non-Gaussian line noise. In this work, we study the use of CNNs for all-sky searches when data are contaminated by line noise. We trained our CNN to classify data into four classes: only Gaussian noise, astrophysical signals injected into Gaussian noise, line noise and Gaussian noise, and an astrophysical signal contaminated by line noise and Gaussian noise. Our CNN safely discriminates the presence and the absence of line noise. In the absence of a line noise, the CNN gives a false alarm probability of 0.5\% and can detect an astrophysical signal with the amplitude of $\log_{10} \hat{h}_0 \gtrsim -1.64$ with 95\% detection probability. On the other hand, if line noise exists in the data, the CNN's false alarm probability increases compared to the case in which line noise is absent. To remedy it, we tried to modify the detection criterion.
The sensitivity depth when a line is present is estimated as $\mathcal{D}^{\mathrm{95\%}} \simeq 3.62$, with the false alarm probability of 10\%.
In terms of the computational time of this pipeline, the preprocess requires $O(10^{19})$ floating-point operations. It is more efficient than the standard methods, though we put the difference of the parameter range and the strain duration aside.
Also, the estimated computational time for candidate selection by the CNN is $O(10^8)$ sec with a single GPU. Improving the hardware and using multiple GPUs would enable us to use CNNs in a real search. 

Accounting for the conditions we neglected is necessary to apply our method to real data. In this work, we ignored the non-stationarity of detector noise, the gaps in the strain data, and the use of multiple detectors. As for line noise, we did not treat the finite coherence time and the comb-like pattern. Also, we need to simulate CGWs with larger $\dot{f}$, and train CNNs to specifically handle this case. We have shown that our CNN is sensitive to astrophysical signals with $|\dot{f}| \lesssim 10^{-12}$ Hz/sec even if it is trained with monochromatic waveforms. On the other hand, standard all-sky search pipelines are sensitive to a signal with $|\dot{f}|\lesssim 10^{-8}$ Hz/sec. Considering the effect of $\dot{f}$ would also be useful to discriminate a line from an astrophysical signal because they have different frequency evolutions. We will extend our method to handle signals with $|\dot{f}| \gtrsim 10^{-12}$ Hz/sec in the future.

Our method includes various parameters to be optimized. The duration of an SFT segment, $\Sub{T}{seg}$, is one of the crucial parameters governing the sensitivity. If we choose a short $\Sub{T}{seg}$, the frequency resolution becomes coarse, leading to signal power being contained in one frequency bin even for large $\dot{f}$. At the same time, line noise will also stay within a frequency bin. Thus, the confusion with line noise could be serious. On the other hand, if we use a long $\Sub{T}{seg}$, the confusion with a line noise will be suppressed; however, the range of $\dot{f}$ in which our method can be applied will become even more limited. To manage the trade-off, we need to try our method with various values of $\Sub{T}{seg}$. 

Another parameter is the residual phase $\delta\Phi_\ast$. If it is small, the signal after the preprocessing step can become large, resulting in better sensitivity. But, the number of the grid points in the sky, and therefore the computational cost, also increase. We should determine $\delta\Phi_\ast$ by considering the trade-off between the computational cost and the sensitivity. Optimizing these parameters will be done in future works.

Our systematic studies of the efficiency of CNNs to detect (quasi-)monochromatic CGWs in the presence of line noise are the first of their kind. They represent a significant step forward towards better understanding and applying CNNs in an real all-sky search. As we show, CNNs can be used to greatly reduce the computational cost compared to existing all-sky search methods, while maintaining impressive sensitivity towards CGW signals in both the presence and absence of line noise.
\begin{acknowledgements}
We thank Chris Messenger for carefully checking our draft and giving us valuable comments. We also thank Marco Cavaglia and Rodrigo Tenorio for giving us useful comments.

This work was supported by JSPS KAKENHI Grant Number JP17H06358 (and also JP17H06357), \textit{A01: Testing gravity theories using gravitational waves}, as a part of the innovative research area, ``Gravitational wave physics and astronomy: Genesis'', and JP20K03928. A.L.M. is a beneficiary of a FSR Incoming Post-doctoral Fellowship.

This material is based upon work supported by NSF's
LIGO Laboratory which is a major facility fully funded
by the National Science Foundation.
\end{acknowledgements}

\appendix
\section{Neural network}
\label{sec:neural network}

\begin{figure}[t]
    \centering
    \includegraphics[width=5cm]{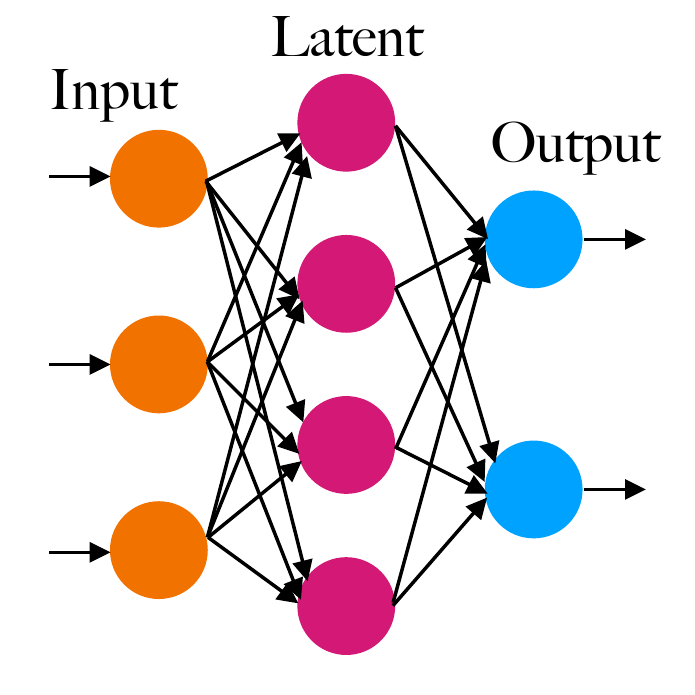}
    \caption{Schematic picture of an artificial neural network. It consists of two fully-connected layers.}
    \label{fig:neural network}
\end{figure}

An \textit{artificial neural network} (ANN) is widely used in  big-data analysis, e.g., image recognition, natural language processing (see~\cite{Goodfellow-et-al-2016} as a textbook). An elementary unit of an ANN is called a \textit{neuron} that is inspired by neural cells in human brains. A neuron can take several values as inputs from other neurons, carry out a linear transformation, and return outputs after a nonlinear transformation called an \textit{activation function}. A number of neurons are stacked into a \textit{layer}, and an ANN has several layers stacked. The input data are fed into the first layer, and the output of the first layer is passed to the second layer, and so on. As a whole, the input data flows through an ANN to the last layer (the \textit{output layer}). Usually, the information goes through in one direction from input to output, called \textit{forward calculation}.
Each layer transforms an input vector $\ve{x} \in \mathbb{R}^{\Sub{N}{in}}$ to an output vector $\ve{o} \in \mathbb{R}^{\Sub{N}{out}}$ by the transformation defined by
\begin{equation}
  z_i = \sum_{j=1}^{\Sub{N}{in}} w_{ij} x_j + b_i\,,\qquad
  o_i = g(z_i)\,.
  \label{eq: fully connected layer}
\end{equation}
Here, $w_{ij}$ and $b_i$ are called \textit{weight} and \textit{bias}, respectively. The function $g$ is an activation function. In this work, we use \textit{rectified linear unit (ReLU)}~\cite{Nair_relu} defined by
\begin{equation}
  g(x) = \begin{cases}
    x & x \geq 0 \\
    0 & x < 0\,.
  \end{cases}
\end{equation}
The transformation given by Eq.~\eqref{eq: fully connected layer} is often named a \textit{fully-connected layer} because all elements of an input vector affect every element of an output vector. It can be schematically pictured by the neurons connected by directed arrows (see Fig.~\ref{fig:neural network}). 
The number of parameters in Eq.~\eqref{eq: fully connected layer} is determined by 
\begin{equation}
    \Sub{N}{in} \times \Sub{N}{out} + \Sub{N}{out}\,,
    \label{eq:tunable params fully connected layer}
\end{equation}
where the first and second term corresponds to the size of the weight and the bias, respectively.

As stated previously, all elements in an input vector $\ve{x}$ connect to every element in an output vector $\ve{z}$ in a fully-connected layer. Fukushima~\cite{Fukushima1980} proposed \textit{Neocognitron} that has a structure in which each element of an output vector connects to a local portion of the input vector. It can be written as
\begin{equation}
    z_i = \sum_{j=1}^{K} w_j x_{s(i-1)+j-1} + b_i\,.
    \label{eq:convolution}
\end{equation}
Here, $K$ is the length of the region where a neuron in a next layer sees at a time. The weight $w$ is often referred to as a \textit{filter}, and $K$ is the size of the filter. In the transformation~\eqref{eq:convolution}, a small region of the input vector is convolved with a filter. The filter is gradually shifted by the width of $s$ to cover the input vector. LeCun~\cite{lecun-89} shows that the connection given by Eq.~\eqref{eq:convolution} is advantageous for extracting local patterns characterizing the input vector. Nowadays, the structure given by Eq.~\eqref{eq:convolution} is referred to as \textit{convolutional layer} and is widely applied especially to image recognition tasks. Another property of convolutional layers is \textit{weight sharing}. In a fully-connected layer~\eqref{eq: fully connected layer}, the number of weights contained in a layer is given by multiplication of the input dimension and the output dimension. It can become a tremendous number of parameters and easily stall the training. Sharing the weights between every output element can significantly reduce the number of tunable parameters and make training an ANN faster.

An input vector of a convolutional layer can be a two-dimensional tensor denoted by $x_{ai} \in \mathbb{R}^{\Sub{C}{in} \times \Sub{N}{in}}$. The index $i$ shows an array of the data. Another index $a$ represents a \textit{channel} which corresponds to the different types of input data. For example, a color image can be characterized by three integers corresponding to the primary colors, i.e., red, green, and blue. A color picture can be represented by three datasets that have the same size of the picture and whose values determine each color's strength. The number of channels of an input is denoted by $\Sub{C}{in}$. The weights can also be shared among different channels. Taking into account the channels, we can write the process carried out in a convolutional layer as
\begin{equation}
  z_{ai} = \sum_{b=1}^{\Sub{C}{in}} \sum_{j=1}^K w_{abj} x_{b,s(i-1)+j-1} + b_{a}\,.
  \label{eq: convolutional layer}
\end{equation}
The weights in a convolutional layer can be represented by three-dimensional tensors, $w_{abj} \in \mathbb{R}^{\Sub{C}{in} \times \Sub{C}{out} \times K}$. The bias $b_a \in \mathbb{R}^{\Sub{C}{out}}$ is a constant for each channel. The number of parameters can be obtained by
\begin{equation}
    \Sub{C}{in} \times \Sub{C}{out} \times K + \Sub{C}{out}\,.
    \label{eq: tunable parmas conv}
\end{equation}

We explained that, by virtue of the weight sharing, a convolutional layer is cheaper than a fully-connected layer in terms of the number of tunable parameters. There is another way to contract the number of data points, which is called \textit{pooling}. Similarly to convolutional layers, pooling reduces the size of an input vector with a particular transformation. In this work, we use a \textit{max pooling} defined by
\begin{equation}
  z_{ai} = \max_{j=1,2,\cdots,K} x_{a,K(i-1)+j}\,.
  \label{eq: maxpooling layer}
\end{equation}

Typically, convolutional layers and pooling layers extract the essential features of the input data. After that, the following fully-connected layers exploit the extracted features to give predictions. A neural network having convolutional layers is often called a \textit{convolutional neural network} (CNN).

In our work, we apply a CNN to detect the CGWs. Detection of astrophysical signals is a typical classification problem where the classifier predicts the class to which a given data likely belongs. In the beginning, all tunable parameters in the neural network are initialized by assigning random values. In this state, the neural network cannot give any reliable predictions. Therefore, we need to tune all weights of the neural network (\textit{training}). Here we can generate simulated training data from the signal and noise models described in Sec.~\ref{sec:waveform and preprocess}. Each simulation can be labeled by a class based on the model (e.g., Gaussian noise only, astrophysical signal injected into Gaussian noise). In other words, we a priori know the class where data should be classified. In general, the training with a data set containing pairs of an input and a target value is named the \textit{supervised learning}. In supervised learning, the neural network is trained so that it can accurately reproduce the target data corresponding to the input data.

The output of the neural network should be appropriate for the problem we try to solve. For classification, the \textit{softmax} transformation defined by
\begin{equation}
  p_i \coloneqq \frac{\exp[x_i]}{\sum_{j=1}^{\Sub{N}{class}} \exp[x_j]}\,,
  \label{eq: softmax layer}
\end{equation}
is widely used. Here, $\Sub{N}{class}$ is the number of classes. Each element $p_i$ of the output means the probability that a given data belongs to the $i$-th class. The \textit{one-hot representation (1-of-$K$ representation)} is a standard representation of the target vector for the classification problem. A target vector $\ve{t}$ represents a vector living in $\{0,1\}^{N_\mathrm{class}}$. If an input data belongs to the $i$-th classes, only $i$-th components of the target vector takes a value 1. For example, if the data belongs to the 1st class, the target vector is represented by
\begin{equation}
  \ve{t} = (1,0,0,\cdots,0)\,.
  \label{eq: one hot rep}
\end{equation}
The vector represented as Eq.~\eqref{eq: one hot rep} can also be interpreted in terms of the probability. That is, each element of a vector $\ve{t}$ gives a probability that the input data is in each class. For the training data, we know the class to which the data belongs. Therefore, it is reasonable to assign a probability unity for the true class and zero for the rest.

In the training, the neural network's predictions and the target values need to be compared. The use of a \textit{loss function} provides us with a quantitative comparison between the predictions and the targets. Depending on the problem we try to solve, we need to carefully choose a loss function. The \textit{cross entropy loss} gives a distance between two probabilities. For a discrete probability, it is defined by
\begin{equation}
  L(\ve{p},\ve{t}) \coloneqq - \sum_{i=1}^{\Sub{N}{class}} t_i \ln p_i\,.
  \label{eq: cross entropy loss}
\end{equation}
The weights of the neural network are optimized so that the expected value of the loss function over the dataset is minimized. We cannot analytically optimize the weights; therefore, we need an iterative scheme to obtain better weights. Schematically, an optimization scheme can be written as
\begin{equation}
  w' = w - \eta \pdiff{L}{w}\,,
  \label{eq: updating weight}
\end{equation}
where $w$ is a weight in the neural network, $w'$ is an updated value of the weight, and $\eta$ is so-called \textit{learning rate} and governs how sensitive the updated amount is to the gradients of the loss function. During the training, the neural network is gradually optimized by repeating a set of processes (1) to feed input data to the ANN, (2) to return the prediction, (3) to evaluate the loss function and its gradients, (4) to update the weights.

\begin{figure}[t]
    \centering
    \includegraphics[width=8.5cm]{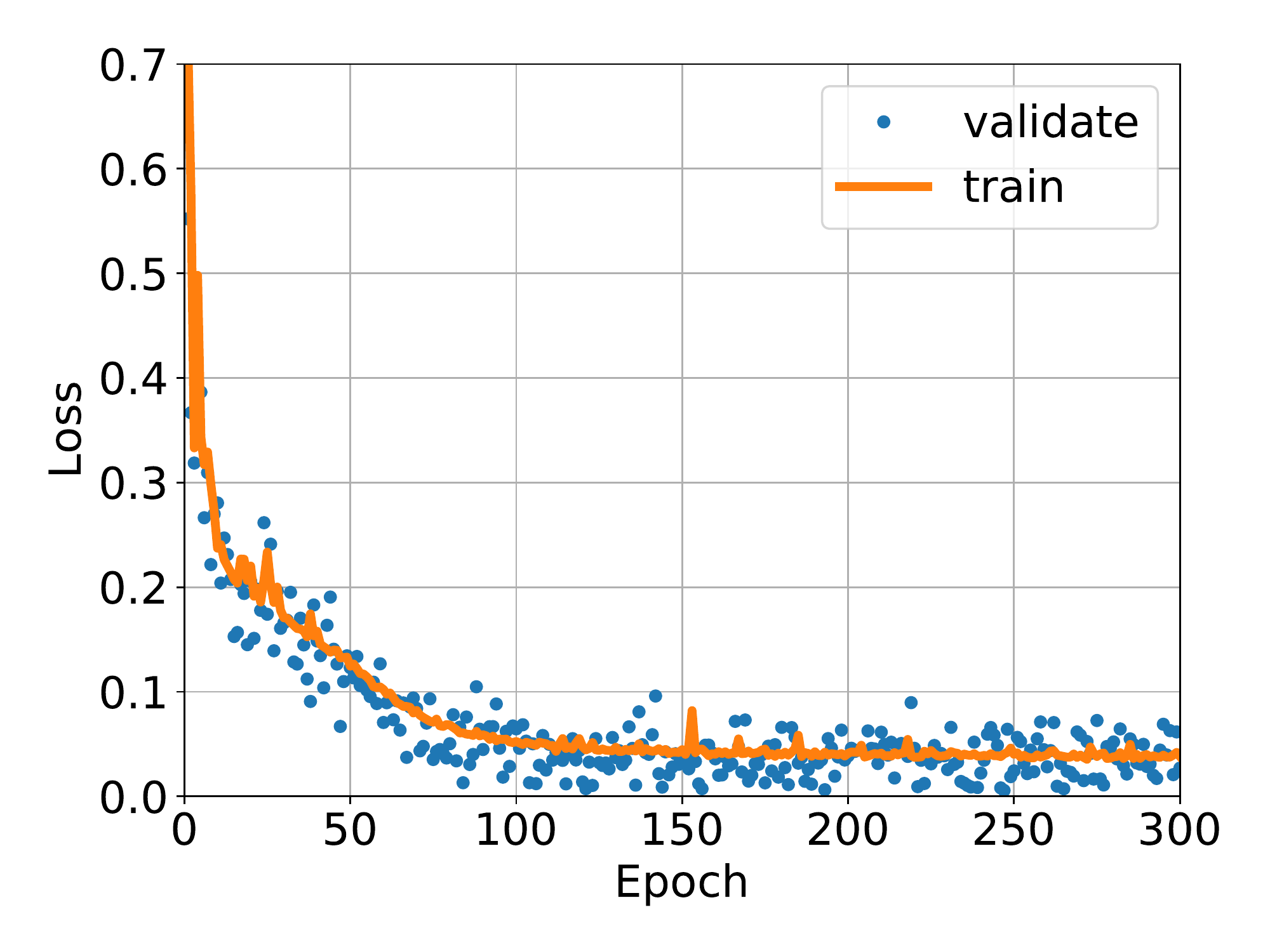}
    \caption{\label{fig:train curve}
    Training curve of CNN of four classes. Orange line shows training loss. Blue dots indicate validate losses. The validation loss have a larger variance than the training loss. This is because of the difference of the number of the training data and the validation data.}
\end{figure}

Due to a large number of tunable parameters, neural networks sometimes learn to only fit the training data. If that happens, neural networks do not have high accuracy on new data. Such a phenomenon is called \textit{overfitting}. To check whether overfitting occurs, we prepare another dataset (\textit{validation data}) and monitor the loss of the validation data in each epoch. If the neural network overfits the training data, the validation loss gradually deviates from the training loss and can increase. Fig.~\ref{fig:train curve} shows the training and validation loss as a function of the training epoch. We find that the loss in our CNN tends to converge well, and the validation loss follows the training loss. Thus, overfitting did not happen during training, and we use the last state of the CNN for testing.

\section{Fine-tuning of CNN}
\label{sec:finetuning}
In general, it takes much time to train a neural network from scratch. Therefore, as a first step, a neural network is trained for a more manageable problem than the one we try to solve finally. This step is called \textit{pretraining}. Then, a pretrained neural network is optimized for the problem we want to solve. The technique optimizing a neural network in a hierarchical manner is called \textit{fine-tuning}. In this appendix, our CNN is fine-tuned, including $\dot{f}$.

\begin{figure}[t]
    \centering
    \includegraphics[width=8cm]{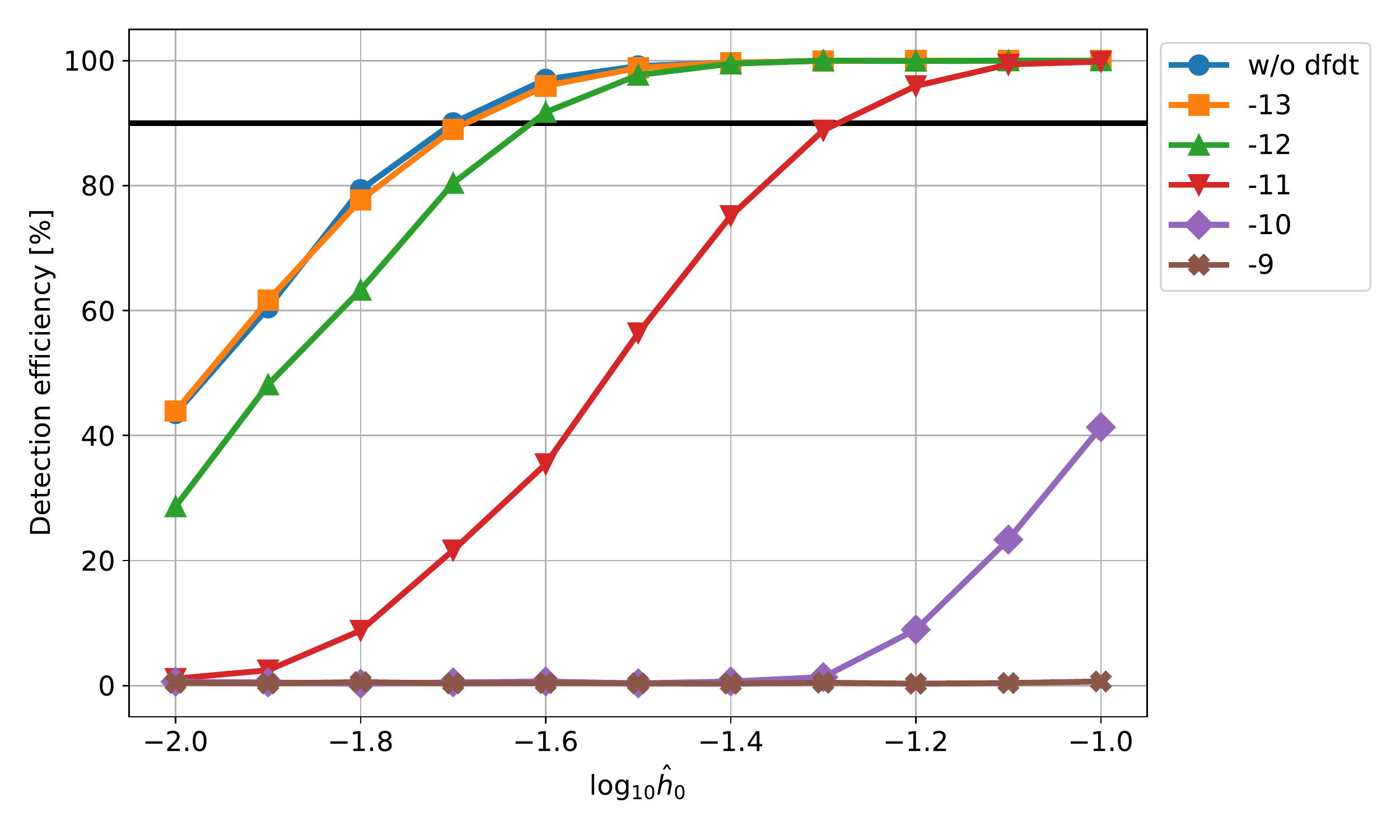}
    \caption{Detection probabilities for astrophysical signals with various values of $\dot{f}$. We set $\dot{f}$ from $-1.0\times 10^{-13}$ to $-1.0\times10^{-9}$ Hz/sec. }
    \label{fig:detection probability ft4cnn dfdt}
\end{figure}

As we explained, for $|\dot{f}| \gtrsim 10^{-11} \mathrm{Hz/sec}$, the signal power would dissipate to several frequency bins. Therefore, we guess that it is useless to include the signal with $|\dot{f}| \gtrsim 10^{-11} \mathrm{Hz/sec}$ in the training data. The data set is generated with the same set up as shown in Sec.~\ref{sec: method} except that $\dot{f}$ is randomly sampled from $10^{-14} \leq |\dot{f}| \leq 10^{-11} \mathrm{Hz/sec}$. We use the trained CNN as an initial state, set the learning rate to $10^{-4}$ and update the weights of the fully-connected layers for 150 epochs with the frozen convolutional layers. Figure~\ref{fig:detection probability ft4cnn dfdt} shows the detection probabilities for signals with various $\dot{f}$. For $|\dot{f}|\leq 10^{-11}$ Hz/sec, the detection probabilities get improved by the fine-tune. It does not improved for $|\dot{f}| = 10^{-9}$ Hz/sec, but it is a predictable result because the dataset for fine-tuning does not include the signal with $\dot{f} = 10^{9}$ Hz/sec.
%

\bibliography{references}

\begin{thebibliography}{67}%
\makeatletter
\providecommand \@ifxundefined [1]{%
 \@ifx{#1\undefined}
}%
\providecommand \@ifnum [1]{%
 \ifnum #1\expandafter \@firstoftwo
 \else \expandafter \@secondoftwo
 \fi
}%
\providecommand \@ifx [1]{%
 \ifx #1\expandafter \@firstoftwo
 \else \expandafter \@secondoftwo
 \fi
}%
\providecommand \natexlab [1]{#1}%
\providecommand \enquote  [1]{``#1''}%
\providecommand \bibnamefont  [1]{#1}%
\providecommand \bibfnamefont [1]{#1}%
\providecommand \citenamefont [1]{#1}%
\providecommand \href@noop [0]{\@secondoftwo}%
\providecommand \href [0]{\begingroup \@sanitize@url \@href}%
\providecommand \@href[1]{\@@startlink{#1}\@@href}%
\providecommand \@@href[1]{\endgroup#1\@@endlink}%
\providecommand \@sanitize@url [0]{\catcode `\\12\catcode `\$12\catcode
  `\&12\catcode `\#12\catcode `\^12\catcode `\_12\catcode `\%12\relax}%
\providecommand \@@startlink[1]{}%
\providecommand \@@endlink[0]{}%
\providecommand \url  [0]{\begingroup\@sanitize@url \@url }%
\providecommand \@url [1]{\endgroup\@href {#1}{\urlprefix }}%
\providecommand \urlprefix  [0]{URL }%
\providecommand \Eprint [0]{\href }%
\providecommand \doibase [0]{https://doi.org/}%
\providecommand \selectlanguage [0]{\@gobble}%
\providecommand \bibinfo  [0]{\@secondoftwo}%
\providecommand \bibfield  [0]{\@secondoftwo}%
\providecommand \translation [1]{[#1]}%
\providecommand \BibitemOpen [0]{}%
\providecommand \bibitemStop [0]{}%
\providecommand \bibitemNoStop [0]{.\EOS\space}%
\providecommand \EOS [0]{\spacefactor3000\relax}%
\providecommand \BibitemShut  [1]{\csname bibitem#1\endcsname}%
\let\auto@bib@innerbib\@empty
\bibitem [{\citenamefont {Lasky}(2015)}]{Lasky:2015uia}%
  \BibitemOpen
  \bibfield  {author} {\bibinfo {author} {\bibfnamefont {P.~D.}\ \bibnamefont
  {Lasky}},\ }\href {https://doi.org/10.1017/pasa.2015.35} {\bibfield
  {journal} {\bibinfo  {journal} {Publ. Astron. Soc. Austral.}\ }\textbf
  {\bibinfo {volume} {32}},\ \bibinfo {pages} {e034} (\bibinfo {year}
  {2015})},\ \Eprint {https://arxiv.org/abs/1508.06643} {arXiv:1508.06643
  [astro-ph.HE]} \BibitemShut {NoStop}%
\bibitem [{\citenamefont {Glampedakis}\ and\ \citenamefont
  {Gualtieri}(2018)}]{Glampedakis:2017nqy}%
  \BibitemOpen
  \bibfield  {author} {\bibinfo {author} {\bibfnamefont {K.}~\bibnamefont
  {Glampedakis}}\ and\ \bibinfo {author} {\bibfnamefont {L.}~\bibnamefont
  {Gualtieri}},\ }\href {https://doi.org/10.1007/978-3-319-97616-7_12}
  {\bibfield  {journal} {\bibinfo  {journal} {Astrophys. Space Sci. Libr.}\
  }\textbf {\bibinfo {volume} {457}},\ \bibinfo {pages} {673} (\bibinfo {year}
  {2018})},\ \Eprint {https://arxiv.org/abs/1709.07049} {arXiv:1709.07049
  [astro-ph.HE]} \BibitemShut {NoStop}%
\bibitem [{\citenamefont {Riles}(2017)}]{riles2017recent}%
  \BibitemOpen
  \bibfield  {author} {\bibinfo {author} {\bibfnamefont {K.}~\bibnamefont
  {Riles}},\ }\href@noop {} {\bibfield  {journal} {\bibinfo  {journal} {Modern
  Physics Letters A}\ }\textbf {\bibinfo {volume} {32}},\ \bibinfo {pages}
  {1730035} (\bibinfo {year} {2017})}\BibitemShut {NoStop}%
\bibitem [{\citenamefont {Sieniawska}\ and\ \citenamefont
  {Bejger}(2019)}]{Sieniawska:2019hmd}%
  \BibitemOpen
  \bibfield  {author} {\bibinfo {author} {\bibfnamefont {M.}~\bibnamefont
  {Sieniawska}}\ and\ \bibinfo {author} {\bibfnamefont {M.}~\bibnamefont
  {Bejger}},\ }\href {https://doi.org/10.3390/universe5110217} {\bibfield
  {journal} {\bibinfo  {journal} {Universe}\ }\textbf {\bibinfo {volume} {5}},\
  \bibinfo {pages} {217} (\bibinfo {year} {2019})},\ \Eprint
  {https://arxiv.org/abs/1909.12600} {arXiv:1909.12600 [astro-ph.HE]}
  \BibitemShut {NoStop}%
\bibitem [{\citenamefont {Tenorio}\ \emph {et~al.}(2021)\citenamefont
  {Tenorio}, \citenamefont {Keitel},\ and\ \citenamefont
  {Sintes}}]{Tenorio:2021wmz}%
  \BibitemOpen
  \bibfield  {author} {\bibinfo {author} {\bibfnamefont {R.}~\bibnamefont
  {Tenorio}}, \bibinfo {author} {\bibfnamefont {D.}~\bibnamefont {Keitel}},\
  and\ \bibinfo {author} {\bibfnamefont {A.~M.}\ \bibnamefont {Sintes}},\
  }\href {https://doi.org/10.3390/universe7120474} {\bibfield  {journal}
  {\bibinfo  {journal} {Universe}\ }\textbf {\bibinfo {volume} {7}},\ \bibinfo
  {pages} {474} (\bibinfo {year} {2021})},\ \Eprint
  {https://arxiv.org/abs/2111.12575} {arXiv:2111.12575 [gr-qc]} \BibitemShut
  {NoStop}%
\bibitem [{\citenamefont {{Lorimer}}\ and\ \citenamefont
  {{Kramer}}(2004)}]{2004hpa..book.....L}%
  \BibitemOpen
  \bibfield  {author} {\bibinfo {author} {\bibfnamefont {D.~R.}\ \bibnamefont
  {{Lorimer}}}\ and\ \bibinfo {author} {\bibfnamefont {M.}~\bibnamefont
  {{Kramer}}},\ }\href@noop {} {\emph {\bibinfo {title} {{Handbook of Pulsar
  Astronomy}}}},\ Vol.~\bibinfo {volume} {4}\ (\bibinfo {year}
  {2004})\BibitemShut {NoStop}%
\bibitem [{Note1()}]{Note1}%
  \BibitemOpen
  \bibinfo {note} {\protect \url
  {http://www.atnf.csiro.au/people/pulsar/psrcat/}}\BibitemShut {NoStop}%
\bibitem [{\citenamefont {Manchester}\ \emph {et~al.}(2005)\citenamefont
  {Manchester}, \citenamefont {Hobbs}, \citenamefont {Teoh},\ and\
  \citenamefont {Hobbs}}]{Manchester:2004bp}%
  \BibitemOpen
  \bibfield  {author} {\bibinfo {author} {\bibfnamefont {R.~N.}\ \bibnamefont
  {Manchester}}, \bibinfo {author} {\bibfnamefont {G.~B.}\ \bibnamefont
  {Hobbs}}, \bibinfo {author} {\bibfnamefont {A.}~\bibnamefont {Teoh}},\ and\
  \bibinfo {author} {\bibfnamefont {M.}~\bibnamefont {Hobbs}},\ }\href
  {https://doi.org/10.1086/428488} {\bibfield  {journal} {\bibinfo  {journal}
  {Astron. J.}\ }\textbf {\bibinfo {volume} {129}},\ \bibinfo {pages} {1993}
  (\bibinfo {year} {2005})},\ \Eprint {https://arxiv.org/abs/astro-ph/0412641}
  {arXiv:astro-ph/0412641} \BibitemShut {NoStop}%
\bibitem [{\citenamefont {Arvanitaki}\ \emph {et~al.}(2010)\citenamefont
  {Arvanitaki}, \citenamefont {Dimopoulos}, \citenamefont {Dubovsky},
  \citenamefont {Kaloper},\ and\ \citenamefont
  {March-Russell}}]{Arvanitaki:2009fg}%
  \BibitemOpen
  \bibfield  {author} {\bibinfo {author} {\bibfnamefont {A.}~\bibnamefont
  {Arvanitaki}}, \bibinfo {author} {\bibfnamefont {S.}~\bibnamefont
  {Dimopoulos}}, \bibinfo {author} {\bibfnamefont {S.}~\bibnamefont
  {Dubovsky}}, \bibinfo {author} {\bibfnamefont {N.}~\bibnamefont {Kaloper}},\
  and\ \bibinfo {author} {\bibfnamefont {J.}~\bibnamefont {March-Russell}},\
  }\href {https://doi.org/10.1103/PhysRevD.81.123530} {\bibfield  {journal}
  {\bibinfo  {journal} {Phys. Rev. D}\ }\textbf {\bibinfo {volume} {81}},\
  \bibinfo {pages} {123530} (\bibinfo {year} {2010})},\ \Eprint
  {https://arxiv.org/abs/0905.4720} {arXiv:0905.4720 [hep-th]} \BibitemShut
  {NoStop}%
\bibitem [{\citenamefont {Brito}\ \emph {et~al.}(2015)\citenamefont {Brito},
  \citenamefont {Cardoso},\ and\ \citenamefont {Pani}}]{Brito:2015oca}%
  \BibitemOpen
  \bibfield  {author} {\bibinfo {author} {\bibfnamefont {R.}~\bibnamefont
  {Brito}}, \bibinfo {author} {\bibfnamefont {V.}~\bibnamefont {Cardoso}},\
  and\ \bibinfo {author} {\bibfnamefont {P.}~\bibnamefont {Pani}},\ }\href
  {https://doi.org/10.1007/978-3-319-19000-6} {\bibfield  {journal} {\bibinfo
  {journal} {Lect. Notes Phys.}\ }\textbf {\bibinfo {volume} {906}},\ \bibinfo
  {pages} {pp.1} (\bibinfo {year} {2015})},\ \Eprint
  {https://arxiv.org/abs/1501.06570} {arXiv:1501.06570 [gr-qc]} \BibitemShut
  {NoStop}%
\bibitem [{\citenamefont {D'Antonio}\ \emph {et~al.}(2018)\citenamefont
  {D'Antonio} \emph {et~al.}}]{DAntonio:2018sff}%
  \BibitemOpen
  \bibfield  {author} {\bibinfo {author} {\bibfnamefont {S.}~\bibnamefont
  {D'Antonio}} \emph {et~al.},\ }\href
  {https://doi.org/10.1103/PhysRevD.98.103017} {\bibfield  {journal} {\bibinfo
  {journal} {Phys. Rev. D}\ }\textbf {\bibinfo {volume} {98}},\ \bibinfo
  {pages} {103017} (\bibinfo {year} {2018})},\ \Eprint
  {https://arxiv.org/abs/1809.07202} {arXiv:1809.07202 [gr-qc]} \BibitemShut
  {NoStop}%
\bibitem [{\citenamefont {Isi}\ \emph {et~al.}(2019)\citenamefont {Isi},
  \citenamefont {Sun}, \citenamefont {Brito},\ and\ \citenamefont
  {Melatos}}]{isi2019directed}%
  \BibitemOpen
  \bibfield  {author} {\bibinfo {author} {\bibfnamefont {M.}~\bibnamefont
  {Isi}}, \bibinfo {author} {\bibfnamefont {L.}~\bibnamefont {Sun}}, \bibinfo
  {author} {\bibfnamefont {R.}~\bibnamefont {Brito}},\ and\ \bibinfo {author}
  {\bibfnamefont {A.}~\bibnamefont {Melatos}},\ }\href@noop {} {\bibfield
  {journal} {\bibinfo  {journal} {Physical Review D}\ }\textbf {\bibinfo
  {volume} {99}},\ \bibinfo {pages} {084042} (\bibinfo {year}
  {2019})}\BibitemShut {NoStop}%
\bibitem [{\citenamefont {Palomba}\ \emph {et~al.}(2019)\citenamefont {Palomba}
  \emph {et~al.}}]{Palomba:2019vxe}%
  \BibitemOpen
  \bibfield  {author} {\bibinfo {author} {\bibfnamefont {C.}~\bibnamefont
  {Palomba}} \emph {et~al.},\ }\href
  {https://doi.org/10.1103/PhysRevLett.123.171101} {\bibfield  {journal}
  {\bibinfo  {journal} {Phys. Rev. Lett.}\ }\textbf {\bibinfo {volume} {123}},\
  \bibinfo {pages} {171101} (\bibinfo {year} {2019})},\ \Eprint
  {https://arxiv.org/abs/1909.08854} {arXiv:1909.08854 [astro-ph.HE]}
  \BibitemShut {NoStop}%
\bibitem [{\citenamefont {Miller}\ \emph {et~al.}(2021)\citenamefont {Miller},
  \citenamefont {Clesse}, \citenamefont {De~Lillo}, \citenamefont {Bruno},
  \citenamefont {Depasse},\ and\ \citenamefont {Tanasijczuk}}]{Miller:2020kmv}%
  \BibitemOpen
  \bibfield  {author} {\bibinfo {author} {\bibfnamefont {A.~L.}\ \bibnamefont
  {Miller}}, \bibinfo {author} {\bibfnamefont {S.}~\bibnamefont {Clesse}},
  \bibinfo {author} {\bibfnamefont {F.}~\bibnamefont {De~Lillo}}, \bibinfo
  {author} {\bibfnamefont {G.}~\bibnamefont {Bruno}}, \bibinfo {author}
  {\bibfnamefont {A.}~\bibnamefont {Depasse}},\ and\ \bibinfo {author}
  {\bibfnamefont {A.}~\bibnamefont {Tanasijczuk}},\ }\href
  {https://doi.org/10.1016/j.dark.2021.100836} {\bibfield  {journal} {\bibinfo
  {journal} {Phys. Dark Univ.}\ }\textbf {\bibinfo {volume} {32}},\ \bibinfo
  {pages} {100836} (\bibinfo {year} {2021})},\ \Eprint
  {https://arxiv.org/abs/2012.12983} {arXiv:2012.12983 [astro-ph.HE]}
  \BibitemShut {NoStop}%
\bibitem [{\citenamefont {Miller}\ \emph {et~al.}(2022)\citenamefont {Miller},
  \citenamefont {Aggarwal}, \citenamefont {Clesse},\ and\ \citenamefont
  {De~Lillo}}]{Miller:2021knj}%
  \BibitemOpen
  \bibfield  {author} {\bibinfo {author} {\bibfnamefont {A.~L.}\ \bibnamefont
  {Miller}}, \bibinfo {author} {\bibfnamefont {N.}~\bibnamefont {Aggarwal}},
  \bibinfo {author} {\bibfnamefont {S.}~\bibnamefont {Clesse}},\ and\ \bibinfo
  {author} {\bibfnamefont {F.}~\bibnamefont {De~Lillo}},\ }\href
  {https://doi.org/10.1103/PhysRevD.105.062008} {\bibfield  {journal} {\bibinfo
   {journal} {Phys. Rev. D}\ }\textbf {\bibinfo {volume} {105}},\ \bibinfo
  {pages} {062008} (\bibinfo {year} {2022})},\ \Eprint
  {https://arxiv.org/abs/2110.06188} {arXiv:2110.06188 [gr-qc]} \BibitemShut
  {NoStop}%
\bibitem [{\citenamefont {Pujolas}\ \emph {et~al.}(2021)\citenamefont
  {Pujolas}, \citenamefont {Vaskonen},\ and\ \citenamefont
  {Veerm\"ae}}]{Pujolas:2021yaw}%
  \BibitemOpen
  \bibfield  {author} {\bibinfo {author} {\bibfnamefont {O.}~\bibnamefont
  {Pujolas}}, \bibinfo {author} {\bibfnamefont {V.}~\bibnamefont {Vaskonen}},\
  and\ \bibinfo {author} {\bibfnamefont {H.}~\bibnamefont {Veerm\"ae}},\ }\href
  {https://doi.org/10.1103/PhysRevD.104.083521} {\bibfield  {journal} {\bibinfo
   {journal} {Phys. Rev. D}\ }\textbf {\bibinfo {volume} {104}},\ \bibinfo
  {pages} {083521} (\bibinfo {year} {2021})},\ \Eprint
  {https://arxiv.org/abs/2107.03379} {arXiv:2107.03379 [astro-ph.CO]}
  \BibitemShut {NoStop}%
\bibitem [{\citenamefont {Guo}\ and\ \citenamefont
  {Miller}(2022)}]{Guo:2022sdd}%
  \BibitemOpen
  \bibfield  {author} {\bibinfo {author} {\bibfnamefont {H.}~\bibnamefont
  {Guo}}\ and\ \bibinfo {author} {\bibfnamefont {A.}~\bibnamefont {Miller}},\
  }\href@noop {} {\  (\bibinfo {year} {2022})},\ \Eprint
  {https://arxiv.org/abs/2205.10359} {arXiv:2205.10359 [astro-ph.IM]}
  \BibitemShut {NoStop}%
\bibitem [{\citenamefont {Jaranowski}\ \emph {et~al.}(1998)\citenamefont
  {Jaranowski}, \citenamefont {Krolak},\ and\ \citenamefont
  {Schutz}}]{Jaranowski:1998qm}%
  \BibitemOpen
  \bibfield  {author} {\bibinfo {author} {\bibfnamefont {P.}~\bibnamefont
  {Jaranowski}}, \bibinfo {author} {\bibfnamefont {A.}~\bibnamefont {Krolak}},\
  and\ \bibinfo {author} {\bibfnamefont {B.~F.}\ \bibnamefont {Schutz}},\
  }\href {https://doi.org/10.1103/PhysRevD.58.063001} {\bibfield  {journal}
  {\bibinfo  {journal} {Phys. Rev. D}\ }\textbf {\bibinfo {volume} {58}},\
  \bibinfo {pages} {063001} (\bibinfo {year} {1998})},\ \Eprint
  {https://arxiv.org/abs/gr-qc/9804014} {arXiv:gr-qc/9804014} \BibitemShut
  {NoStop}%
\bibitem [{\citenamefont {Astone}\ \emph {et~al.}(2014)\citenamefont {Astone},
  \citenamefont {Colla}, \citenamefont {D'Antonio}, \citenamefont {Frasca},\
  and\ \citenamefont {Palomba}}]{Astone:2014esa}%
  \BibitemOpen
  \bibfield  {author} {\bibinfo {author} {\bibfnamefont {P.}~\bibnamefont
  {Astone}}, \bibinfo {author} {\bibfnamefont {A.}~\bibnamefont {Colla}},
  \bibinfo {author} {\bibfnamefont {S.}~\bibnamefont {D'Antonio}}, \bibinfo
  {author} {\bibfnamefont {S.}~\bibnamefont {Frasca}},\ and\ \bibinfo {author}
  {\bibfnamefont {C.}~\bibnamefont {Palomba}},\ }\href
  {https://doi.org/10.1103/PhysRevD.90.042002} {\bibfield  {journal} {\bibinfo
  {journal} {Phys. Rev. D}\ }\textbf {\bibinfo {volume} {90}},\ \bibinfo
  {pages} {042002} (\bibinfo {year} {2014})},\ \Eprint
  {https://arxiv.org/abs/1407.8333} {arXiv:1407.8333 [astro-ph.IM]}
  \BibitemShut {NoStop}%
\bibitem [{\citenamefont {Krishnan}\ \emph {et~al.}(2004)\citenamefont
  {Krishnan}, \citenamefont {Sintes}, \citenamefont {Papa}, \citenamefont
  {Schutz}, \citenamefont {Frasca},\ and\ \citenamefont
  {Palomba}}]{Krishnan:2004sv}%
  \BibitemOpen
  \bibfield  {author} {\bibinfo {author} {\bibfnamefont {B.}~\bibnamefont
  {Krishnan}}, \bibinfo {author} {\bibfnamefont {A.~M.}\ \bibnamefont
  {Sintes}}, \bibinfo {author} {\bibfnamefont {M.~A.}\ \bibnamefont {Papa}},
  \bibinfo {author} {\bibfnamefont {B.~F.}\ \bibnamefont {Schutz}}, \bibinfo
  {author} {\bibfnamefont {S.}~\bibnamefont {Frasca}},\ and\ \bibinfo {author}
  {\bibfnamefont {C.}~\bibnamefont {Palomba}},\ }\href
  {https://doi.org/10.1103/PhysRevD.70.082001} {\bibfield  {journal} {\bibinfo
  {journal} {Phys. Rev. D}\ }\textbf {\bibinfo {volume} {70}},\ \bibinfo
  {pages} {082001} (\bibinfo {year} {2004})},\ \Eprint
  {https://arxiv.org/abs/gr-qc/0407001} {arXiv:gr-qc/0407001} \BibitemShut
  {NoStop}%
\bibitem [{\citenamefont {Bayley}\ \emph {et~al.}(2019)\citenamefont {Bayley},
  \citenamefont {Woan},\ and\ \citenamefont {Messenger}}]{Bayley:2019bcb}%
  \BibitemOpen
  \bibfield  {author} {\bibinfo {author} {\bibfnamefont {J.}~\bibnamefont
  {Bayley}}, \bibinfo {author} {\bibfnamefont {G.}~\bibnamefont {Woan}},\ and\
  \bibinfo {author} {\bibfnamefont {C.}~\bibnamefont {Messenger}},\ }\href
  {https://doi.org/10.1103/PhysRevD.100.023006} {\bibfield  {journal} {\bibinfo
   {journal} {Phys. Rev. D}\ }\textbf {\bibinfo {volume} {100}},\ \bibinfo
  {pages} {023006} (\bibinfo {year} {2019})},\ \Eprint
  {https://arxiv.org/abs/1903.12614} {arXiv:1903.12614 [astro-ph.IM]}
  \BibitemShut {NoStop}%
\bibitem [{\citenamefont {Abbott}\ \emph
  {et~al.}(2017{\natexlab{a}})\citenamefont {Abbott} \emph
  {et~al.}}]{LIGOScientific:2017csd}%
  \BibitemOpen
  \bibfield  {author} {\bibinfo {author} {\bibfnamefont {B.~P.}\ \bibnamefont
  {Abbott}} \emph {et~al.} (\bibinfo {collaboration} {LIGO Scientific,
  Virgo}),\ }\href {https://doi.org/10.1103/PhysRevD.96.062002} {\bibfield
  {journal} {\bibinfo  {journal} {Phys. Rev. D}\ }\textbf {\bibinfo {volume}
  {96}},\ \bibinfo {pages} {062002} (\bibinfo {year} {2017}{\natexlab{a}})},\
  \Eprint {https://arxiv.org/abs/1707.02667} {arXiv:1707.02667 [gr-qc]}
  \BibitemShut {NoStop}%
\bibitem [{\citenamefont {Abbott}\ \emph
  {et~al.}(2017{\natexlab{b}})\citenamefont {Abbott} \emph
  {et~al.}}]{LIGOScientific:2017wva}%
  \BibitemOpen
  \bibfield  {author} {\bibinfo {author} {\bibfnamefont {B.~P.}\ \bibnamefont
  {Abbott}} \emph {et~al.} (\bibinfo {collaboration} {LIGO Scientific,
  Virgo}),\ }\href {https://doi.org/10.1103/PhysRevD.96.122004} {\bibfield
  {journal} {\bibinfo  {journal} {Phys. Rev. D}\ }\textbf {\bibinfo {volume}
  {96}},\ \bibinfo {pages} {122004} (\bibinfo {year} {2017}{\natexlab{b}})},\
  \Eprint {https://arxiv.org/abs/1707.02669} {arXiv:1707.02669 [gr-qc]}
  \BibitemShut {NoStop}%
\bibitem [{\citenamefont {Abbott}\ \emph {et~al.}(2018)\citenamefont {Abbott}
  \emph {et~al.}}]{LIGOScientific:2018gpj}%
  \BibitemOpen
  \bibfield  {author} {\bibinfo {author} {\bibfnamefont {B.~P.}\ \bibnamefont
  {Abbott}} \emph {et~al.} (\bibinfo {collaboration} {LIGO Scientific,
  Virgo}),\ }\href {https://doi.org/10.1103/PhysRevD.97.102003} {\bibfield
  {journal} {\bibinfo  {journal} {Phys. Rev. D}\ }\textbf {\bibinfo {volume}
  {97}},\ \bibinfo {pages} {102003} (\bibinfo {year} {2018})},\ \Eprint
  {https://arxiv.org/abs/1802.05241} {arXiv:1802.05241 [gr-qc]} \BibitemShut
  {NoStop}%
\bibitem [{\citenamefont {Abbott}\ \emph {et~al.}(2019)\citenamefont {Abbott}
  \emph {et~al.}}]{LIGOScientific:2019yhl}%
  \BibitemOpen
  \bibfield  {author} {\bibinfo {author} {\bibfnamefont {B.~P.}\ \bibnamefont
  {Abbott}} \emph {et~al.} (\bibinfo {collaboration} {LIGO Scientific,
  Virgo}),\ }\href {https://doi.org/10.1103/PhysRevD.100.024004} {\bibfield
  {journal} {\bibinfo  {journal} {Phys. Rev. D}\ }\textbf {\bibinfo {volume}
  {100}},\ \bibinfo {pages} {024004} (\bibinfo {year} {2019})},\ \Eprint
  {https://arxiv.org/abs/1903.01901} {arXiv:1903.01901 [astro-ph.HE]}
  \BibitemShut {NoStop}%
\bibitem [{\citenamefont {Abbott}\ \emph
  {et~al.}(2021{\natexlab{a}})\citenamefont {Abbott} \emph
  {et~al.}}]{LIGOScientific:2020qhb}%
  \BibitemOpen
  \bibfield  {author} {\bibinfo {author} {\bibfnamefont {R.}~\bibnamefont
  {Abbott}} \emph {et~al.} (\bibinfo {collaboration} {LIGO Scientific,
  Virgo}),\ }\href {https://doi.org/10.1103/PhysRevD.103.064017} {\bibfield
  {journal} {\bibinfo  {journal} {Phys. Rev. D}\ }\textbf {\bibinfo {volume}
  {103}},\ \bibinfo {pages} {064017} (\bibinfo {year} {2021}{\natexlab{a}})},\
  \Eprint {https://arxiv.org/abs/2012.12128} {arXiv:2012.12128 [gr-qc]}
  \BibitemShut {NoStop}%
\bibitem [{\citenamefont {Abbott}\ \emph
  {et~al.}(2021{\natexlab{b}})\citenamefont {Abbott} \emph
  {et~al.}}]{LIGOScientific:2021inr}%
  \BibitemOpen
  \bibfield  {author} {\bibinfo {author} {\bibfnamefont {R.}~\bibnamefont
  {Abbott}} \emph {et~al.} (\bibinfo {collaboration} {LIGO Scientific,
  VIRGO}),\ }\Eprint {https://arxiv.org/abs/2111.15116} {arXiv:2111.15116
  [gr-qc]}  (\bibinfo {year} {2021}{\natexlab{b}})\BibitemShut {NoStop}%
\bibitem [{\citenamefont {Abbott}\ \emph
  {et~al.}(2021{\natexlab{c}})\citenamefont {Abbott} \emph
  {et~al.}}]{LIGOScientific:2021jlr}%
  \BibitemOpen
  \bibfield  {author} {\bibinfo {author} {\bibfnamefont {R.}~\bibnamefont
  {Abbott}} \emph {et~al.} (\bibinfo {collaboration} {LIGO Scientific, VIRGO,
  KAGRA}),\ }\Eprint {https://arxiv.org/abs/2111.15507} {arXiv:2111.15507
  [astro-ph.HE]}  (\bibinfo {year} {2021}{\natexlab{c}})\BibitemShut {NoStop}%
\bibitem [{\citenamefont {Abbott}\ \emph
  {et~al.}(2021{\natexlab{d}})\citenamefont {Abbott} \emph
  {et~al.}}]{LIGOScientific:2021quq}%
  \BibitemOpen
  \bibfield  {author} {\bibinfo {author} {\bibfnamefont {R.}~\bibnamefont
  {Abbott}} \emph {et~al.} (\bibinfo {collaboration} {LIGO Scientific, VIRGO,
  KAGRA}),\ }\Eprint {https://arxiv.org/abs/2112.10990} {arXiv:2112.10990
  [gr-qc]}  (\bibinfo {year} {2021}{\natexlab{d}})\BibitemShut {NoStop}%
\bibitem [{\citenamefont {Abbott}\ \emph {et~al.}(2022)\citenamefont {Abbott}
  \emph {et~al.}}]{LIGOScientific:2022pjk}%
  \BibitemOpen
  \bibfield  {author} {\bibinfo {author} {\bibfnamefont {R.}~\bibnamefont
  {Abbott}} \emph {et~al.} (\bibinfo {collaboration} {LIGO Scientific, VIRGO,
  KAGRA}),\ }\Eprint {https://arxiv.org/abs/2201.00697} {arXiv:2201.00697
  [gr-qc]}  (\bibinfo {year} {2022})\BibitemShut {NoStop}%
\bibitem [{\citenamefont {Covas}\ \emph {et~al.}(2018)\citenamefont {Covas}
  \emph {et~al.}}]{LSC:2018vzm}%
  \BibitemOpen
  \bibfield  {author} {\bibinfo {author} {\bibfnamefont {P.~B.}\ \bibnamefont
  {Covas}} \emph {et~al.} (\bibinfo {collaboration} {LSC}),\ }\href
  {https://doi.org/10.1103/PhysRevD.97.082002} {\bibfield  {journal} {\bibinfo
  {journal} {Phys. Rev. D}\ }\textbf {\bibinfo {volume} {97}},\ \bibinfo
  {pages} {082002} (\bibinfo {year} {2018})},\ \Eprint
  {https://arxiv.org/abs/1801.07204} {arXiv:1801.07204 [astro-ph.IM]}
  \BibitemShut {NoStop}%
\bibitem [{\citenamefont {Cuoco}\ \emph {et~al.}(2021)\citenamefont {Cuoco}
  \emph {et~al.}}]{Cuoco:2020ogp}%
  \BibitemOpen
  \bibfield  {author} {\bibinfo {author} {\bibfnamefont {E.}~\bibnamefont
  {Cuoco}} \emph {et~al.},\ }\href {https://doi.org/10.1088/2632-2153/abb93a}
  {\bibfield  {journal} {\bibinfo  {journal} {Mach. Learn. Sci. Tech.}\
  }\textbf {\bibinfo {volume} {2}},\ \bibinfo {pages} {011002} (\bibinfo {year}
  {2021})},\ \Eprint {https://arxiv.org/abs/2005.03745} {arXiv:2005.03745
  [astro-ph.HE]} \BibitemShut {NoStop}%
\bibitem [{\citenamefont {George}\ and\ \citenamefont
  {Huerta}(2018{\natexlab{a}})}]{George:2016hay}%
  \BibitemOpen
  \bibfield  {author} {\bibinfo {author} {\bibfnamefont {D.}~\bibnamefont
  {George}}\ and\ \bibinfo {author} {\bibfnamefont {E.~A.}\ \bibnamefont
  {Huerta}},\ }\href {https://doi.org/10.1103/PhysRevD.97.044039} {\bibfield
  {journal} {\bibinfo  {journal} {Phys. Rev. D}\ }\textbf {\bibinfo {volume}
  {97}},\ \bibinfo {pages} {044039} (\bibinfo {year} {2018}{\natexlab{a}})},\
  \Eprint {https://arxiv.org/abs/1701.00008} {arXiv:1701.00008 [astro-ph.IM]}
  \BibitemShut {NoStop}%
\bibitem [{\citenamefont {George}\ and\ \citenamefont
  {Huerta}(2018{\natexlab{b}})}]{George:2017pmj}%
  \BibitemOpen
  \bibfield  {author} {\bibinfo {author} {\bibfnamefont {D.}~\bibnamefont
  {George}}\ and\ \bibinfo {author} {\bibfnamefont {E.~A.}\ \bibnamefont
  {Huerta}},\ }\href {https://doi.org/10.1016/j.physletb.2017.12.053}
  {\bibfield  {journal} {\bibinfo  {journal} {Phys. Lett. B}\ }\textbf
  {\bibinfo {volume} {778}},\ \bibinfo {pages} {64} (\bibinfo {year}
  {2018}{\natexlab{b}})},\ \Eprint {https://arxiv.org/abs/1711.03121}
  {arXiv:1711.03121 [gr-qc]} \BibitemShut {NoStop}%
\bibitem [{\citenamefont {Colgan}\ \emph {et~al.}(2020)\citenamefont {Colgan},
  \citenamefont {Corley}, \citenamefont {Lau}, \citenamefont {Bartos},
  \citenamefont {Wright}, \citenamefont {Marka},\ and\ \citenamefont
  {Marka}}]{Colgan:2019lyo}%
  \BibitemOpen
  \bibfield  {author} {\bibinfo {author} {\bibfnamefont {R.~E.}\ \bibnamefont
  {Colgan}}, \bibinfo {author} {\bibfnamefont {K.~R.}\ \bibnamefont {Corley}},
  \bibinfo {author} {\bibfnamefont {Y.}~\bibnamefont {Lau}}, \bibinfo {author}
  {\bibfnamefont {I.}~\bibnamefont {Bartos}}, \bibinfo {author} {\bibfnamefont
  {J.~N.}\ \bibnamefont {Wright}}, \bibinfo {author} {\bibfnamefont
  {Z.}~\bibnamefont {Marka}},\ and\ \bibinfo {author} {\bibfnamefont
  {S.}~\bibnamefont {Marka}},\ }\href
  {https://doi.org/10.1103/PhysRevD.101.102003} {\bibfield  {journal} {\bibinfo
   {journal} {Phys. Rev. D}\ }\textbf {\bibinfo {volume} {101}},\ \bibinfo
  {pages} {102003} (\bibinfo {year} {2020})},\ \Eprint
  {https://arxiv.org/abs/1911.11831} {arXiv:1911.11831 [astro-ph.IM]}
  \BibitemShut {NoStop}%
\bibitem [{\citenamefont {Schmidt}\ \emph {et~al.}(2021)\citenamefont
  {Schmidt}, \citenamefont {Breschi}, \citenamefont {Gamba}, \citenamefont
  {Pagano}, \citenamefont {Rettegno}, \citenamefont {Riemenschneider},
  \citenamefont {Bernuzzi}, \citenamefont {Nagar},\ and\ \citenamefont
  {Del~Pozzo}}]{Schmidt:2020yuu}%
  \BibitemOpen
  \bibfield  {author} {\bibinfo {author} {\bibfnamefont {S.}~\bibnamefont
  {Schmidt}}, \bibinfo {author} {\bibfnamefont {M.}~\bibnamefont {Breschi}},
  \bibinfo {author} {\bibfnamefont {R.}~\bibnamefont {Gamba}}, \bibinfo
  {author} {\bibfnamefont {G.}~\bibnamefont {Pagano}}, \bibinfo {author}
  {\bibfnamefont {P.}~\bibnamefont {Rettegno}}, \bibinfo {author}
  {\bibfnamefont {G.}~\bibnamefont {Riemenschneider}}, \bibinfo {author}
  {\bibfnamefont {S.}~\bibnamefont {Bernuzzi}}, \bibinfo {author}
  {\bibfnamefont {A.}~\bibnamefont {Nagar}},\ and\ \bibinfo {author}
  {\bibfnamefont {W.}~\bibnamefont {Del~Pozzo}},\ }\href
  {https://doi.org/10.1103/PhysRevD.103.043020} {\bibfield  {journal} {\bibinfo
   {journal} {Phys. Rev. D}\ }\textbf {\bibinfo {volume} {103}},\ \bibinfo
  {pages} {043020} (\bibinfo {year} {2021})},\ \Eprint
  {https://arxiv.org/abs/2011.01958} {arXiv:2011.01958 [gr-qc]} \BibitemShut
  {NoStop}%
\bibitem [{\citenamefont {Gabbard}\ \emph {et~al.}(2022)\citenamefont
  {Gabbard}, \citenamefont {Messenger}, \citenamefont {Heng}, \citenamefont
  {Tonolini},\ and\ \citenamefont {Murray-Smith}}]{Gabbard:2019rde}%
  \BibitemOpen
  \bibfield  {author} {\bibinfo {author} {\bibfnamefont {H.}~\bibnamefont
  {Gabbard}}, \bibinfo {author} {\bibfnamefont {C.}~\bibnamefont {Messenger}},
  \bibinfo {author} {\bibfnamefont {I.~S.}\ \bibnamefont {Heng}}, \bibinfo
  {author} {\bibfnamefont {F.}~\bibnamefont {Tonolini}},\ and\ \bibinfo
  {author} {\bibfnamefont {R.}~\bibnamefont {Murray-Smith}},\ }\href
  {https://doi.org/10.1038/s41567-021-01425-7} {\bibfield  {journal} {\bibinfo
  {journal} {Nature Phys.}\ }\textbf {\bibinfo {volume} {18}},\ \bibinfo
  {pages} {112} (\bibinfo {year} {2022})},\ \Eprint
  {https://arxiv.org/abs/1909.06296} {arXiv:1909.06296 [astro-ph.IM]}
  \BibitemShut {NoStop}%
\bibitem [{\citenamefont {Chua}\ and\ \citenamefont
  {Vallisneri}(2020)}]{Chua:2019wwt}%
  \BibitemOpen
  \bibfield  {author} {\bibinfo {author} {\bibfnamefont {A.~J.~K.}\
  \bibnamefont {Chua}}\ and\ \bibinfo {author} {\bibfnamefont {M.}~\bibnamefont
  {Vallisneri}},\ }\href {https://doi.org/10.1103/PhysRevLett.124.041102}
  {\bibfield  {journal} {\bibinfo  {journal} {Phys. Rev. Lett.}\ }\textbf
  {\bibinfo {volume} {124}},\ \bibinfo {pages} {041102} (\bibinfo {year}
  {2020})},\ \Eprint {https://arxiv.org/abs/1909.05966} {arXiv:1909.05966
  [gr-qc]} \BibitemShut {NoStop}%
\bibitem [{\citenamefont {Green}\ \emph {et~al.}(2020)\citenamefont {Green},
  \citenamefont {Simpson},\ and\ \citenamefont {Gair}}]{Green:2020hst}%
  \BibitemOpen
  \bibfield  {author} {\bibinfo {author} {\bibfnamefont {S.~R.}\ \bibnamefont
  {Green}}, \bibinfo {author} {\bibfnamefont {C.}~\bibnamefont {Simpson}},\
  and\ \bibinfo {author} {\bibfnamefont {J.}~\bibnamefont {Gair}},\ }\href
  {https://doi.org/10.1103/PhysRevD.102.104057} {\bibfield  {journal} {\bibinfo
   {journal} {Phys. Rev. D}\ }\textbf {\bibinfo {volume} {102}},\ \bibinfo
  {pages} {104057} (\bibinfo {year} {2020})},\ \Eprint
  {https://arxiv.org/abs/2002.07656} {arXiv:2002.07656 [astro-ph.IM]}
  \BibitemShut {NoStop}%
\bibitem [{\citenamefont {Dax}\ \emph {et~al.}(2021)\citenamefont {Dax},
  \citenamefont {Green}, \citenamefont {Gair}, \citenamefont {Macke},
  \citenamefont {Buonanno},\ and\ \citenamefont {Sch\"olkopf}}]{Dax:2021tsq}%
  \BibitemOpen
  \bibfield  {author} {\bibinfo {author} {\bibfnamefont {M.}~\bibnamefont
  {Dax}}, \bibinfo {author} {\bibfnamefont {S.~R.}\ \bibnamefont {Green}},
  \bibinfo {author} {\bibfnamefont {J.}~\bibnamefont {Gair}}, \bibinfo {author}
  {\bibfnamefont {J.~H.}\ \bibnamefont {Macke}}, \bibinfo {author}
  {\bibfnamefont {A.}~\bibnamefont {Buonanno}},\ and\ \bibinfo {author}
  {\bibfnamefont {B.}~\bibnamefont {Sch\"olkopf}},\ }\href
  {https://doi.org/10.1103/PhysRevLett.127.241103} {\bibfield  {journal}
  {\bibinfo  {journal} {Phys. Rev. Lett.}\ }\textbf {\bibinfo {volume} {127}},\
  \bibinfo {pages} {241103} (\bibinfo {year} {2021})},\ \Eprint
  {https://arxiv.org/abs/2106.12594} {arXiv:2106.12594 [gr-qc]} \BibitemShut
  {NoStop}%
\bibitem [{\citenamefont {Kuo}\ and\ \citenamefont {Lin}(2022)}]{Kuo:2021qtt}%
  \BibitemOpen
  \bibfield  {author} {\bibinfo {author} {\bibfnamefont {H.-S.}\ \bibnamefont
  {Kuo}}\ and\ \bibinfo {author} {\bibfnamefont {F.-L.}\ \bibnamefont {Lin}},\
  }\href {https://doi.org/10.1103/PhysRevD.105.044016} {\bibfield  {journal}
  {\bibinfo  {journal} {Phys. Rev. D}\ }\textbf {\bibinfo {volume} {105}},\
  \bibinfo {pages} {044016} (\bibinfo {year} {2022})},\ \Eprint
  {https://arxiv.org/abs/2107.10730} {arXiv:2107.10730 [gr-qc]} \BibitemShut
  {NoStop}%
\bibitem [{\citenamefont {Nakano}\ \emph {et~al.}(2019)\citenamefont {Nakano},
  \citenamefont {Narikawa}, \citenamefont {Oohara}, \citenamefont {Sakai},
  \citenamefont {Shinkai}, \citenamefont {Takahashi}, \citenamefont {Tanaka},
  \citenamefont {Uchikata}, \citenamefont {Yamamoto},\ and\ \citenamefont
  {Yamamoto}}]{Nakano:2018vay}%
  \BibitemOpen
  \bibfield  {author} {\bibinfo {author} {\bibfnamefont {H.}~\bibnamefont
  {Nakano}}, \bibinfo {author} {\bibfnamefont {T.}~\bibnamefont {Narikawa}},
  \bibinfo {author} {\bibfnamefont {K.-i.}\ \bibnamefont {Oohara}}, \bibinfo
  {author} {\bibfnamefont {K.}~\bibnamefont {Sakai}}, \bibinfo {author}
  {\bibfnamefont {H.-a.}\ \bibnamefont {Shinkai}}, \bibinfo {author}
  {\bibfnamefont {H.}~\bibnamefont {Takahashi}}, \bibinfo {author}
  {\bibfnamefont {T.}~\bibnamefont {Tanaka}}, \bibinfo {author} {\bibfnamefont
  {N.}~\bibnamefont {Uchikata}}, \bibinfo {author} {\bibfnamefont
  {S.}~\bibnamefont {Yamamoto}},\ and\ \bibinfo {author} {\bibfnamefont
  {T.~S.}\ \bibnamefont {Yamamoto}},\ }\href
  {https://doi.org/10.1103/PhysRevD.99.124032} {\bibfield  {journal} {\bibinfo
  {journal} {Phys. Rev. D}\ }\textbf {\bibinfo {volume} {99}},\ \bibinfo
  {pages} {124032} (\bibinfo {year} {2019})},\ \Eprint
  {https://arxiv.org/abs/1811.06443} {arXiv:1811.06443 [gr-qc]} \BibitemShut
  {NoStop}%
\bibitem [{\citenamefont {Shen}\ \emph {et~al.}(2022)\citenamefont {Shen},
  \citenamefont {Huerta}, \citenamefont {O'Shea}, \citenamefont {Kumar},\ and\
  \citenamefont {Zhao}}]{Shen:2019vep}%
  \BibitemOpen
  \bibfield  {author} {\bibinfo {author} {\bibfnamefont {H.}~\bibnamefont
  {Shen}}, \bibinfo {author} {\bibfnamefont {E.~A.}\ \bibnamefont {Huerta}},
  \bibinfo {author} {\bibfnamefont {E.}~\bibnamefont {O'Shea}}, \bibinfo
  {author} {\bibfnamefont {P.}~\bibnamefont {Kumar}},\ and\ \bibinfo {author}
  {\bibfnamefont {Z.}~\bibnamefont {Zhao}},\ }\href
  {https://doi.org/10.1088/2632-2153/ac3843} {\bibfield  {journal} {\bibinfo
  {journal} {Mach. Learn. Sci. Tech.}\ }\textbf {\bibinfo {volume} {3}},\
  \bibinfo {pages} {015007} (\bibinfo {year} {2022})},\ \Eprint
  {https://arxiv.org/abs/1903.01998} {arXiv:1903.01998 [gr-qc]} \BibitemShut
  {NoStop}%
\bibitem [{\citenamefont {Yamamoto}\ and\ \citenamefont
  {Tanaka}(2020)}]{Yamamoto:2020rse}%
  \BibitemOpen
  \bibfield  {author} {\bibinfo {author} {\bibfnamefont {T.~S.}\ \bibnamefont
  {Yamamoto}}\ and\ \bibinfo {author} {\bibfnamefont {T.}~\bibnamefont
  {Tanaka}},\ }\Eprint {https://arxiv.org/abs/2002.12095} {arXiv:2002.12095
  [gr-qc]}  (\bibinfo {year} {2020})\BibitemShut {NoStop}%
\bibitem [{\citenamefont {Bhagwat}\ and\ \citenamefont
  {Pacilio}(2021)}]{Bhagwat:2021kfa}%
  \BibitemOpen
  \bibfield  {author} {\bibinfo {author} {\bibfnamefont {S.}~\bibnamefont
  {Bhagwat}}\ and\ \bibinfo {author} {\bibfnamefont {C.}~\bibnamefont
  {Pacilio}},\ }\href {https://doi.org/10.1103/PhysRevD.104.024030} {\bibfield
  {journal} {\bibinfo  {journal} {Phys. Rev. D}\ }\textbf {\bibinfo {volume}
  {104}},\ \bibinfo {pages} {024030} (\bibinfo {year} {2021})},\ \Eprint
  {https://arxiv.org/abs/2101.07817} {arXiv:2101.07817 [gr-qc]} \BibitemShut
  {NoStop}%
\bibitem [{\citenamefont {Miller}\ \emph {et~al.}(2019)\citenamefont {Miller}
  \emph {et~al.}}]{Miller:2019jtp}%
  \BibitemOpen
  \bibfield  {author} {\bibinfo {author} {\bibfnamefont {A.~L.}\ \bibnamefont
  {Miller}} \emph {et~al.},\ }\href
  {https://doi.org/10.1103/PhysRevD.100.062005} {\bibfield  {journal} {\bibinfo
   {journal} {Phys. Rev. D}\ }\textbf {\bibinfo {volume} {100}},\ \bibinfo
  {pages} {062005} (\bibinfo {year} {2019})},\ \Eprint
  {https://arxiv.org/abs/1909.02262} {arXiv:1909.02262 [astro-ph.IM]}
  \BibitemShut {NoStop}%
\bibitem [{\citenamefont {Miller}\ \emph {et~al.}(2018)\citenamefont {Miller}
  \emph {et~al.}}]{Miller:2018rbg}%
  \BibitemOpen
  \bibfield  {author} {\bibinfo {author} {\bibfnamefont {A.}~\bibnamefont
  {Miller}} \emph {et~al.},\ }\href
  {https://doi.org/10.1103/PhysRevD.98.102004} {\bibfield  {journal} {\bibinfo
  {journal} {Phys. Rev. D}\ }\textbf {\bibinfo {volume} {98}},\ \bibinfo
  {pages} {102004} (\bibinfo {year} {2018})},\ \Eprint
  {https://arxiv.org/abs/1810.09784} {arXiv:1810.09784 [astro-ph.IM]}
  \BibitemShut {NoStop}%
\bibitem [{\citenamefont {Morr\'as}\ \emph {et~al.}(2022)\citenamefont
  {Morr\'as}, \citenamefont {Garc\'\i{}a-Bellido},\ and\ \citenamefont
  {Nesseris}}]{Morras:2021atg}%
  \BibitemOpen
  \bibfield  {author} {\bibinfo {author} {\bibfnamefont {G.}~\bibnamefont
  {Morr\'as}}, \bibinfo {author} {\bibfnamefont {J.}~\bibnamefont
  {Garc\'\i{}a-Bellido}},\ and\ \bibinfo {author} {\bibfnamefont
  {S.}~\bibnamefont {Nesseris}},\ }\href
  {https://doi.org/10.1016/j.dark.2021.100932} {\bibfield  {journal} {\bibinfo
  {journal} {Phys. Dark Univ.}\ }\textbf {\bibinfo {volume} {35}},\ \bibinfo
  {pages} {100932} (\bibinfo {year} {2022})},\ \Eprint
  {https://arxiv.org/abs/2110.08000} {arXiv:2110.08000 [astro-ph.HE]}
  \BibitemShut {NoStop}%
\bibitem [{\citenamefont {Chatterjee}\ \emph {et~al.}(2020)\citenamefont
  {Chatterjee}, \citenamefont {Ghosh}, \citenamefont {Brady}, \citenamefont
  {Kapadia}, \citenamefont {Miller}, \citenamefont {Nissanke},\ and\
  \citenamefont {Pannarale}}]{Chatterjee:2019avs}%
  \BibitemOpen
  \bibfield  {author} {\bibinfo {author} {\bibfnamefont {D.}~\bibnamefont
  {Chatterjee}}, \bibinfo {author} {\bibfnamefont {S.}~\bibnamefont {Ghosh}},
  \bibinfo {author} {\bibfnamefont {P.~R.}\ \bibnamefont {Brady}}, \bibinfo
  {author} {\bibfnamefont {S.~J.}\ \bibnamefont {Kapadia}}, \bibinfo {author}
  {\bibfnamefont {A.~L.}\ \bibnamefont {Miller}}, \bibinfo {author}
  {\bibfnamefont {S.}~\bibnamefont {Nissanke}},\ and\ \bibinfo {author}
  {\bibfnamefont {F.}~\bibnamefont {Pannarale}},\ }\href
  {https://doi.org/10.3847/1538-4357/ab8dbe} {\bibfield  {journal} {\bibinfo
  {journal} {Astrophys. J.}\ }\textbf {\bibinfo {volume} {896}},\ \bibinfo
  {pages} {54} (\bibinfo {year} {2020})},\ \Eprint
  {https://arxiv.org/abs/1911.00116} {arXiv:1911.00116 [astro-ph.IM]}
  \BibitemShut {NoStop}%
\bibitem [{\citenamefont {Mytidis}\ \emph {et~al.}(2019)\citenamefont {Mytidis}
  \emph {et~al.}}]{mytidis2015sensitivity}%
  \BibitemOpen
  \bibfield  {author} {\bibinfo {author} {\bibfnamefont {A.}~\bibnamefont
  {Mytidis}} \emph {et~al.},\ }\href@noop {} {\bibfield  {journal} {\bibinfo
  {journal} {Physical Review D}\ }\textbf {\bibinfo {volume} {99}},\ \bibinfo
  {pages} {024024} (\bibinfo {year} {2019})}\BibitemShut {NoStop}%
\bibitem [{\citenamefont {Dreissigacker}\ \emph {et~al.}(2019)\citenamefont
  {Dreissigacker}, \citenamefont {Sharma}, \citenamefont {Messenger},
  \citenamefont {Zhao},\ and\ \citenamefont {Prix}}]{Dreissigacker:2019edy}%
  \BibitemOpen
  \bibfield  {author} {\bibinfo {author} {\bibfnamefont {C.}~\bibnamefont
  {Dreissigacker}}, \bibinfo {author} {\bibfnamefont {R.}~\bibnamefont
  {Sharma}}, \bibinfo {author} {\bibfnamefont {C.}~\bibnamefont {Messenger}},
  \bibinfo {author} {\bibfnamefont {R.}~\bibnamefont {Zhao}},\ and\ \bibinfo
  {author} {\bibfnamefont {R.}~\bibnamefont {Prix}},\ }\href
  {https://doi.org/10.1103/PhysRevD.100.044009} {\bibfield  {journal} {\bibinfo
   {journal} {Phys. Rev. D}\ }\textbf {\bibinfo {volume} {100}},\ \bibinfo
  {pages} {044009} (\bibinfo {year} {2019})},\ \Eprint
  {https://arxiv.org/abs/1904.13291} {arXiv:1904.13291 [gr-qc]} \BibitemShut
  {NoStop}%
\bibitem [{\citenamefont {Morawski}\ \emph {et~al.}(2020)\citenamefont
  {Morawski}, \citenamefont {Bejger},\ and\ \citenamefont
  {Cieciel\textbackslash{}{a}g}}]{Morawski:2019awi}%
  \BibitemOpen
  \bibfield  {author} {\bibinfo {author} {\bibfnamefont {F.}~\bibnamefont
  {Morawski}}, \bibinfo {author} {\bibfnamefont {M.}~\bibnamefont {Bejger}},\
  and\ \bibinfo {author} {\bibfnamefont {P.}~\bibnamefont
  {Cieciel\textbackslash{}{a}g}},\ }\href
  {https://doi.org/10.1088/2632-2153/ab86c7} {\bibfield  {journal} {\bibinfo
  {journal} {Mach. Learn. Sci. Tech.}\ }\textbf {\bibinfo {volume} {1}},\
  \bibinfo {pages} {025016} (\bibinfo {year} {2020})},\ \Eprint
  {https://arxiv.org/abs/1907.06917} {arXiv:1907.06917 [astro-ph.IM]}
  \BibitemShut {NoStop}%
\bibitem [{\citenamefont {Beheshtipour}\ and\ \citenamefont
  {Papa}(2020)}]{Beheshtipour:2020zhb}%
  \BibitemOpen
  \bibfield  {author} {\bibinfo {author} {\bibfnamefont {B.}~\bibnamefont
  {Beheshtipour}}\ and\ \bibinfo {author} {\bibfnamefont {M.~A.}\ \bibnamefont
  {Papa}},\ }\href {https://doi.org/10.1103/PhysRevD.101.064009} {\bibfield
  {journal} {\bibinfo  {journal} {Phys. Rev. D}\ }\textbf {\bibinfo {volume}
  {101}},\ \bibinfo {pages} {064009} (\bibinfo {year} {2020})},\ \Eprint
  {https://arxiv.org/abs/2001.03116} {arXiv:2001.03116 [gr-qc]} \BibitemShut
  {NoStop}%
\bibitem [{\citenamefont {Bayley}\ \emph {et~al.}(2020)\citenamefont {Bayley},
  \citenamefont {Messenger},\ and\ \citenamefont {Woan}}]{Bayley:2020zfa}%
  \BibitemOpen
  \bibfield  {author} {\bibinfo {author} {\bibfnamefont {J.}~\bibnamefont
  {Bayley}}, \bibinfo {author} {\bibfnamefont {C.}~\bibnamefont {Messenger}},\
  and\ \bibinfo {author} {\bibfnamefont {G.}~\bibnamefont {Woan}},\ }\href
  {https://doi.org/10.1103/PhysRevD.102.083024} {\bibfield  {journal} {\bibinfo
   {journal} {Phys. Rev. D}\ }\textbf {\bibinfo {volume} {102}},\ \bibinfo
  {pages} {083024} (\bibinfo {year} {2020})},\ \Eprint
  {https://arxiv.org/abs/2007.08207} {arXiv:2007.08207 [astro-ph.IM]}
  \BibitemShut {NoStop}%
\bibitem [{\citenamefont {Walsh}\ \emph {et~al.}(2016)\citenamefont {Walsh}
  \emph {et~al.}}]{Walsh:2016hyc}%
  \BibitemOpen
  \bibfield  {author} {\bibinfo {author} {\bibfnamefont {S.}~\bibnamefont
  {Walsh}} \emph {et~al.},\ }\href {https://doi.org/10.1103/PhysRevD.94.124010}
  {\bibfield  {journal} {\bibinfo  {journal} {Phys. Rev. D}\ }\textbf {\bibinfo
  {volume} {94}},\ \bibinfo {pages} {124010} (\bibinfo {year} {2016})},\
  \Eprint {https://arxiv.org/abs/1606.00660} {arXiv:1606.00660 [gr-qc]}
  \BibitemShut {NoStop}%
\bibitem [{\citenamefont {Yamamoto}\ and\ \citenamefont
  {Tanaka}(2021)}]{Yamamoto:2020pus}%
  \BibitemOpen
  \bibfield  {author} {\bibinfo {author} {\bibfnamefont {T.~S.}\ \bibnamefont
  {Yamamoto}}\ and\ \bibinfo {author} {\bibfnamefont {T.}~\bibnamefont
  {Tanaka}},\ }\href {https://doi.org/10.1103/PhysRevD.103.084049} {\bibfield
  {journal} {\bibinfo  {journal} {Phys. Rev. D}\ }\textbf {\bibinfo {volume}
  {103}},\ \bibinfo {pages} {084049} (\bibinfo {year} {2021})},\ \Eprint
  {https://arxiv.org/abs/2011.12522} {arXiv:2011.12522 [gr-qc]} \BibitemShut
  {NoStop}%
\bibitem [{\citenamefont {Jaranowski}\ and\ \citenamefont
  {Krolak}(2009)}]{Jaranowski:2009zz}%
  \BibitemOpen
  \bibfield  {author} {\bibinfo {author} {\bibfnamefont {P.}~\bibnamefont
  {Jaranowski}}\ and\ \bibinfo {author} {\bibfnamefont {A.}~\bibnamefont
  {Krolak}},\ }\href {https://doi.org/10.1017/CBO9780511605482} {\emph
  {\bibinfo {title} {{Analysis of gravitational-wave data}}}}\ (\bibinfo
  {publisher} {Cambridge Univ. Press},\ \bibinfo {address} {Cambridge},\
  \bibinfo {year} {2009})\BibitemShut {NoStop}%
\bibitem [{\citenamefont {Aasi}\ \emph {et~al.}(2015)\citenamefont {Aasi} \emph
  {et~al.}}]{LIGOScientific:2014pky}%
  \BibitemOpen
  \bibfield  {author} {\bibinfo {author} {\bibfnamefont {J.}~\bibnamefont
  {Aasi}} \emph {et~al.} (\bibinfo {collaboration} {LIGO Scientific
  Collaboration}),\ }\href {https://doi.org/10.1088/0264-9381/32/7/074001}
  {\bibfield  {journal} {\bibinfo  {journal} {Class. Quant. Grav.}\ }\textbf
  {\bibinfo {volume} {32}},\ \bibinfo {pages} {074001} (\bibinfo {year}
  {2015})},\ \Eprint {https://arxiv.org/abs/1411.4547} {arXiv:1411.4547
  [gr-qc]} \BibitemShut {NoStop}%
\bibitem [{\citenamefont {Goodfellow}\ \emph {et~al.}(2016)\citenamefont
  {Goodfellow}, \citenamefont {Bengio},\ and\ \citenamefont
  {Courville}}]{Goodfellow-et-al-2016}%
  \BibitemOpen
  \bibfield  {author} {\bibinfo {author} {\bibfnamefont {I.}~\bibnamefont
  {Goodfellow}}, \bibinfo {author} {\bibfnamefont {Y.}~\bibnamefont {Bengio}},\
  and\ \bibinfo {author} {\bibfnamefont {A.}~\bibnamefont {Courville}},\
  }\href@noop {} {\emph {\bibinfo {title} {Deep Learning}}}\ (\bibinfo
  {publisher} {MIT Press},\ \bibinfo {year} {2016})\ \bibinfo {note}
  {\url{http://www.deeplearningbook.org}}\BibitemShut {NoStop}%
\bibitem [{\citenamefont {Nakano}\ \emph {et~al.}(2003)\citenamefont {Nakano},
  \citenamefont {Takahashi}, \citenamefont {Tagoshi},\ and\ \citenamefont
  {Sasaki}}]{Nakano:2003ma}%
  \BibitemOpen
  \bibfield  {author} {\bibinfo {author} {\bibfnamefont {H.}~\bibnamefont
  {Nakano}}, \bibinfo {author} {\bibfnamefont {H.}~\bibnamefont {Takahashi}},
  \bibinfo {author} {\bibfnamefont {H.}~\bibnamefont {Tagoshi}},\ and\ \bibinfo
  {author} {\bibfnamefont {M.}~\bibnamefont {Sasaki}},\ }\href
  {https://doi.org/10.1103/PhysRevD.68.102003} {\bibfield  {journal} {\bibinfo
  {journal} {Phys. Rev. D}\ }\textbf {\bibinfo {volume} {68}},\ \bibinfo
  {pages} {102003} (\bibinfo {year} {2003})},\ \Eprint
  {https://arxiv.org/abs/gr-qc/0306082} {arXiv:gr-qc/0306082} \BibitemShut
  {NoStop}%
\bibitem [{Note2()}]{Note2}%
  \BibitemOpen
  \bibinfo {note} {The Tukey window would reduce both the powers of signal and
  noise. Therefore, Eq.~\protect \textup {\hbox {\mathsurround \z@ \protect
  \normalfont (\ignorespaces \ref {eq: noise variance in ell domain}\unskip
  \@@italiccorr )}} overestimates the variance of noise. Because the CGW signal
  and the line noise are generated with considering the Tukey window, the SNR
  of simulated data could be underestimated. In this sense, our estimation of
  detection efficiency is conservative.}\BibitemShut {Stop}%
\bibitem [{\citenamefont {Kingma}\ and\ \citenamefont
  {Ba}(2014)}]{Kingma:2014vow}%
  \BibitemOpen
  \bibfield  {author} {\bibinfo {author} {\bibfnamefont {D.~P.}\ \bibnamefont
  {Kingma}}\ and\ \bibinfo {author} {\bibfnamefont {J.}~\bibnamefont {Ba}},\
  }\Eprint {https://arxiv.org/abs/1412.6980} {arXiv:1412.6980 [cs.LG]}
  (\bibinfo {year} {2014})\BibitemShut {NoStop}%
\bibitem [{\citenamefont {Paszke}\ \emph {et~al.}(2019)\citenamefont {Paszke},
  \citenamefont {Gross}, \citenamefont {Massa}, \citenamefont {Lerer},
  \citenamefont {Bradbury}, \citenamefont {Chanan}, \citenamefont {Killeen},
  \citenamefont {Lin}, \citenamefont {Gimelshein}, \citenamefont {Antiga},
  \citenamefont {Desmaison}, \citenamefont {Köpf}, \citenamefont {Yang},
  \citenamefont {DeVito}, \citenamefont {Raison}, \citenamefont {Tejani},
  \citenamefont {Chilamkurthy}, \citenamefont {Steiner}, \citenamefont {Fang},
  \citenamefont {Bai},\ and\ \citenamefont {Chintala}}]{Paszke2019}%
  \BibitemOpen
  \bibfield  {author} {\bibinfo {author} {\bibfnamefont {A.}~\bibnamefont
  {Paszke}}, \bibinfo {author} {\bibfnamefont {S.}~\bibnamefont {Gross}},
  \bibinfo {author} {\bibfnamefont {F.}~\bibnamefont {Massa}}, \bibinfo
  {author} {\bibfnamefont {A.}~\bibnamefont {Lerer}}, \bibinfo {author}
  {\bibfnamefont {J.}~\bibnamefont {Bradbury}}, \bibinfo {author}
  {\bibfnamefont {G.}~\bibnamefont {Chanan}}, \bibinfo {author} {\bibfnamefont
  {T.}~\bibnamefont {Killeen}}, \bibinfo {author} {\bibfnamefont
  {Z.}~\bibnamefont {Lin}}, \bibinfo {author} {\bibfnamefont {N.}~\bibnamefont
  {Gimelshein}}, \bibinfo {author} {\bibfnamefont {L.}~\bibnamefont {Antiga}},
  \bibinfo {author} {\bibfnamefont {A.}~\bibnamefont {Desmaison}}, \bibinfo
  {author} {\bibfnamefont {A.}~\bibnamefont {Köpf}}, \bibinfo {author}
  {\bibfnamefont {E.}~\bibnamefont {Yang}}, \bibinfo {author} {\bibfnamefont
  {Z.}~\bibnamefont {DeVito}}, \bibinfo {author} {\bibfnamefont
  {M.}~\bibnamefont {Raison}}, \bibinfo {author} {\bibfnamefont
  {A.}~\bibnamefont {Tejani}}, \bibinfo {author} {\bibfnamefont
  {S.}~\bibnamefont {Chilamkurthy}}, \bibinfo {author} {\bibfnamefont
  {B.}~\bibnamefont {Steiner}}, \bibinfo {author} {\bibfnamefont
  {L.}~\bibnamefont {Fang}}, \bibinfo {author} {\bibfnamefont {J.}~\bibnamefont
  {Bai}},\ and\ \bibinfo {author} {\bibfnamefont {S.}~\bibnamefont
  {Chintala}},\ }\href@noop {} {\bibfield  {journal} {\bibinfo  {journal}
  {Advances in neural information processing systems}\ }\textbf {\bibinfo
  {volume} {32}} (\bibinfo {year} {2019})},\ \Eprint
  {https://arxiv.org/abs/1912.01703} {arXiv:1912.01703 [cs.LG]} \BibitemShut
  {NoStop}%
\bibitem [{Note3()}]{Note3}%
  \BibitemOpen
  \bibinfo {note} {Assuming the use of 10 GPUs with twice faster than GTX1080Ti
  that is used in this work, we have computational time $T_\protect \mathrm
  {CNN} \simeq 5.0 \times 10^6 \protect \mathrm {[sec]}$.}\BibitemShut {Stop}%
\bibitem [{\citenamefont {Nair}\ and\ \citenamefont
  {Hinton}(2010)}]{Nair_relu}%
  \BibitemOpen
  \bibfield  {author} {\bibinfo {author} {\bibfnamefont {V.}~\bibnamefont
  {Nair}}\ and\ \bibinfo {author} {\bibfnamefont {G.~E.}\ \bibnamefont
  {Hinton}},\ }in\ \href@noop {} {\emph {\bibinfo {booktitle} {Proceedings of
  the 27th ICML}}}\ (\bibinfo  {publisher} {Omnipress},\ \bibinfo {address}
  {Madison, WI, USA},\ \bibinfo {year} {2010})\ p.\ \bibinfo {pages}
  {807–814}\BibitemShut {NoStop}%
\bibitem [{\citenamefont {Fukushima}(1980)}]{Fukushima1980}%
  \BibitemOpen
  \bibfield  {author} {\bibinfo {author} {\bibfnamefont {K.}~\bibnamefont
  {Fukushima}},\ }\href {https://doi.org/10.1007/BF00344251} {\bibfield
  {journal} {\bibinfo  {journal} {Biological Cybernetics}\ }\textbf {\bibinfo
  {volume} {36}},\ \bibinfo {pages} {193} (\bibinfo {year} {1980})}\BibitemShut
  {NoStop}%
\bibitem [{\citenamefont {LeCun}(1989)}]{lecun-89}%
  \BibitemOpen
  \bibfield  {author} {\bibinfo {author} {\bibfnamefont {Y.}~\bibnamefont
  {LeCun}},\ }in\ \href@noop {} {\emph {\bibinfo {booktitle} {Connectionism in
  Perspective}}},\ \bibinfo {editor} {edited by\ \bibinfo {editor}
  {\bibfnamefont {R.}~\bibnamefont {Pfeifer}}, \bibinfo {editor} {\bibfnamefont
  {Z.}~\bibnamefont {Schreter}}, \bibinfo {editor} {\bibfnamefont
  {F.}~\bibnamefont {Fogelman}},\ and\ \bibinfo {editor} {\bibfnamefont
  {L.}~\bibnamefont {Steels}}}\ (\bibinfo  {publisher} {Elsevier},\ \bibinfo
  {address} {Zurich, Switzerland},\ \bibinfo {year} {1989})\ \bibinfo {note}
  {an extended version was published as a technical report of the University of
  Toronto}\BibitemShut {NoStop}%
\end{thebibliography}%

\end{document}